\documentclass[11pt,titlepage]{report}
\usepackage[utf8]{inputenc}
\usepackage[T1]{fontenc}
\usepackage[british]{babel}
\usepackage{graphicx}
\usepackage[dvipsnames]{xcolor}
\usepackage[T1]{fontenc}
\usepackage{lmodern}
\usepackage{pdflscape}
\usepackage[sf]{titlesec}
\usepackage[square,numbers]{natbib}
\usepackage{tikz}
\usetikzlibrary{automata, shapes.geometric, arrows.meta, decorations.pathmorphing, positioning, calc, shapes}
\tikzset{
  perspective_box/.style = {draw, fill=blue!20, minimum width=4cm, minimum height=1cm, text centered, thick, top color=blue!10, bottom color=blue!50, transform shape, yslant=-0.4},
  ruby_box/.style = {draw, fill=red!20, minimum width=4cm, minimum height=1cm, text centered, thick, top color=red!10, bottom color=red!50, transform shape, yslant=-0.4},
  cpu_box/.style = {draw, fill=green!20, minimum width=4cm, minimum height=1cm, text centered, thick, top color=green!10, bottom color=green!50, transform shape, yslant=-0.4},
  mmu_box/.style = {draw, fill=violet!20, minimum width=4cm, minimum height=1cm, text centered, thick, top color=blue!10, bottom color=violet!50, transform shape, yslant=-0.4},
    icache_box/.style = {draw, fill=olive!30, minimum width=4cm, minimum height=1cm, text centered, thick, top color=lime!20, bottom color=yellow!50, transform shape, yslant=-0.4},
    dcache_box/.style = {draw, fill=red!40, minimum width=4cm, minimum height=1cm, text centered, thick, top color=red!30, bottom color=red!90, transform shape, yslant=-0.4},
    ins_box/.style = {draw, fill=red!40, minimum width=4cm, minimum height=1cm, text centered, thick, top color=orange!30, bottom color=orange!90, transform shape, yslant=-0.4},
  circle_node/.style = {circle, draw, fill=green!20, text centered, thick, minimum size=2cm, top color=blue!10, bottom color=green!50},
  arrow/.style = {draw, -{Stealth}, thick},
}

\tikzset{
  l1cc/.style = {draw, ellipse, fill=cyan!30, minimum width=4cm, minimum height=2cm, text centered, thick},
  l2cc/.style = {draw, ellipse, fill=blue!40, minimum width=4cm, minimum height=2cm, text centered, thick},
  memcontr/.style = {draw, ellipse, fill=teal!30, minimum width=4cm, minimum height=2cm, text centered, thick},
}

\tikzset{
    queuei/.pic={
        \draw[line width=1pt]
        (0,0) -- ++(1.4cm,0) -- ++(0,-1cm) -- ++(-1.4cm,0);
        \foreach \Val in {1,...,3}
        \draw ([xshift=-\Val*10pt]1.4cm,0) -- ++(0,-1cm);
        \node[above] at (1cm,0) {};   
    },
}

\tikzset{
  queueilong/.pic={
    \draw[line width=1pt]
    (0,0) -- ++(5.6cm,0) -- ++(0,-1cm) -- ++(-5.6cm,0);
    \foreach \Val in {1,...,12}
      \draw ([xshift=-\Val*10pt]5.6cm,0) -- ++(0,-1cm);
  },
}
\tikzset{
  gradient_box/.style={
    draw, thick, rounded corners,
    minimum width=3cm, minimum height=1cm,
    text centered, font=\bfseries,
    top color=#1!20, bottom color=#1!60
  }
}

\usepackage{
  marginnote, 
  environ,
  ragged2e,  
  url,        
  listings,   
  hyperref,   
  lipsum,
  booktabs,
  changepage,
  float,
  dingbat
}
\hypersetup{hidelinks}

\usepackage{etoolbox} 

\newcommand{\benchmarkColor}[1]{%
  \ifstrequal{#1}{gapbs}{\def\colorname{red}}{%
    \ifstrequal{#1}{parsec3}{\def\colorname{blue}}{%
      \ifstrequal{#1}{spec2017}{\def\colorname{ForestGreen}}{%
        \def\colorname{black}}}}}%

\newcommand{\profilingFig}[5]{ 
  \benchmarkColor{#1} 
  \begin{figure}[H]
    \centering
    \includegraphics[width=\textwidth]{./figures_profiling_cfg_figs_#1_#2_#3-#4.pdf}
    \caption{\label{fig:#4_#1}#5}
  \end{figure}
}

\newcommand{\profilingFigs}[3]{ 
  \profilingFig{gapbs}{#1}{gapbs}{#2}{#3 for \textcolor{red}{GAPBS} running #1.}
  \profilingFig{parsec3}{#1}{parsec-3.0}{#2}{#3 for \textcolor{blue}{PARSEC-3.0} running #1.}
  \profilingFig{spec2017}{#1}{spec2017}{#2}{#3 for \textcolor{ForestGreen}{SPEC2017} running #1.}
}

\newcommand{\profilingRefs}[1]{Figures \ref{fig:#1_gapbs}, \ref{fig:#1_parsec3} and \ref{fig:#1_spec2017}}

\usepackage[capitalise]{cleveref}
\Crefname{section}{\S}{\S\S}

\usepackage{xcolor}
\newcommand{\cmark}{\textcolor{green!80!black}{\ding{51}}}
\newcommand{\xmark}{\textcolor{red}{\ding{55}}}

\usepackage{enumitem}   
\setlist[itemize]{itemsep=0pt}
\setlist[itemize,1]{label=$\bullet$}
\setlist[itemize,2]{label=$\triangleright$}
\setlist[itemize,3]{label=\RED{AVOID}}

\usepackage[labelfont=sf]{caption}

\newcommand{\specs}{GNU/Linux Ubuntu 22.04.4 LTS,  Linux kernel 6.5.0-28-generic, 12th Generation Intel\textsuperscript{®} Core\textsuperscript{™} i9-12900K (24 cores), x86\_64 ISA, 128 gigabytes of system memory}

\newcommand\myhrulefill[1]{\leavevmode\leaders\hrule height#1\hfill\kern0pt}
\DeclareCaptionFormat{FigFormat}{{\color{black}\myhrulefill{0.5pt}}\\#1#2#3}
\DeclareCaptionFormat{LstFormat}{\textsf{Listing}~\arabic{chapter}.\arabic{listing}:#2#3}
\floatstyle{ruled}
\newfloat{listing}{thp}{lol}[chapter]
\floatname{listing}{Listing}
\renewenvironment{figure*}[1][]{%
  \begin{figure}[#1]%
    }{%
  \end{figure}}

\renewenvironment{table*}[1][]{%
  \begin{table}[#1]%
    }{%
  \end{table}}

\renewenvironment{listing*}[1][]{%
  \begin{listing}[#1]%
    }{%
  \end{listing}}

\lstnewenvironment{Code}[1][style=std]{\lstset{#1}}{}
\lstnewenvironment{Code_Numbered}[1][style=std,numbers=left]{\lstset{#1}}{}

\lstdefinestyle{std}{
  language=java,
  basicstyle=\small\tt\color{black},
  keywordstyle=\small\tt\bfseries,
  numberstyle=\footnotesize\sf\color{black},
  commentstyle=\small\color{black}\it,
  aboveskip=1ex,
  belowskip=1ex,
  tabsize=2,
  columns=fullflexible,
  xleftmargin=1ex,
  resetmargins=true,
  showstringspaces=false,
  morecomment=[l]{//},
  morecomment=[l]{--},
  morecomment=[s]{/*}{*/},
  escapeinside=@@,
  morekeywords={Frobies},
  moredelim=[is][\textit]{___}{___},
  moredelim=[is][\textbf]{__*}{*__},
  numberbychapter=true
}

\usepackage[activate={true,nocompatibility},final,tracking=true,kerning=true,spacing=true,factor=1100,stretch=10,shrink=10]{microtype}

\newcommand{\RED}[1]{\textcolor{red}{#1}}

\usepackage{pifont}

\usepackage[driver=pdftex,paper=a4paper,margin=1in]{geometry}

\clearpage{\pagestyle{empty}\cleardoublepage}

\begin{document}
\pagenumbering{roman}
\title{Anatomy of the gem5 Simulator:\\AtomicSimpleCPU, TimingSimpleCPU, O3CPU, and Their Interaction with the Ruby Memory System \\[1em]\large\texttt{Using gem5 24.0 with x86\_64 ISA}}
\author{\\[5em]Johan Söderström, \\ Department of IT, Uppsala University \\ johan.soderstrom.9461@student.uu.se \\[1em]Yuan Yao, \\ Department of IT, Uppsala University\\yuan.yao@it.uu.se\\[15em]}
\date{\today\\[1em]\textbf{v1.0}}

\maketitle

\vspace*{3cm}
\section*{Abstract}

gem5 is a popular modular-based computer system simulator, widely used in computer architecture research and known for its long simulation time and steep learning curve. This report examines its three major CPU models: the AtomicSimpleCPU (AS CPU), the TimingSimpleCPU (TS CPU), the Out-of-order (O3) CPU, and their interactions with the memory subsystem. We provide a detailed anatomical overview of each CPU's function call-chains and present how gem5 partitions its execution time for each simulated hardware layer.

We perform our analysis using a lightweight profiler built on Linux's \texttt{perf\_event} interface, with user-configurable options to target specific functions and examine their interactions in detail. By profiling each CPU across a wide selection of benchmarks, we identify their software bottlenecks. Our results show that the \emph{Ruby} memory subsystem consistently accounts for the largest share of execution time in the sequential AS and TS CPUs---primarily during the \emph{instruction fetch} stage. In contrast, the O3 CPU spends a relatively smaller fraction of time in \emph{Ruby}, with most of its time devoted to constructing instruction instances and the various pipeline stages of the CPU.

We believe that the anatomical view of each CPU’s execution flow is valuable for educational purposes, as it clearly illustrates the interactions among simulated components. These insights form a foundation for optimizing gem5’s performance---particularly for the AS, TS, and O3 CPUs. Moreover, our framework can be readily applied to analyze other gem5 components or to develop and evaluate new models.

\tableofcontents
\listoffigures
\listoftables

\cleardoublepage
\pagenumbering{arabic}

\chapter{Introduction}
\label{cha:background}
The gem5 simulator \cite{binkert2011gem5, lowe2020gem5} is pervasively used within academia and industry for computer architecture research since it allows for extensive configuration of simulated computer systems\footnote{Available at: \url{https://www.gem5.org/}}. gem5 is also known for its long execution time, during which the user is ignorant about specifically what portions of the simulator are the most time-consuming. The goal of this report is to explain the runtime behaviour of gem5's AtomicSimpleCPU (AS), TimingSimpleCPU (TS), and Out-of-order CPU (O3), and their interaction with the memory subsystem, to review their bottlenecks and the impact of memory and core scaling---all via extensive profiling detailing the execution time of each simulated hardware component. Furthermore, the profiling methodology facilitates for new architecture design development by providing fine-grained statistics on gem5's components.
\section{gem5 overview}
gem5 is a modular open-source simulator. It simulates computer systems and their components, such as cores, memory systems, and entire operating systems (OS). It is pervasively used in computer architecture research to evaluate system performance and can be tailored to specific hardware configurations, thereby allowing system architectures to be assessed by running an application inside the simulator.

Its simulation consists of discrete events---its time units are measured in \emph{tick}s, and each event executed has been scheduled and dequeued from an event-queue. gem5's simulation is entirely driven by its eponymous top-level {\tt simulate} function, which calls the {\tt doSimLoop} to process one \emph{event} at a time from the {\tt EventQueue}. The {\tt EventQueue} handles all pending events and almost all of its runtime is dedicated to the {\tt serviceOne} function, which dequeues and executes the next scheduled event. This will invoke the CPU to process the event, which is completed in some number of CPU cycles.

gem5 can simulate a wide range of systems from simple CPU configurations to complex memory hiearchies running multiple cores. Its modular design divides each component into objects, meaning that each object can be independently configured and swapped---easing the process of adjusting a complete finished system. This means that gem5's source code is object-oriented---enforcing \emph{separation of concerns}, which also leads to extensive routing of function calls throughout the simulator's backend. gem5 is configured via Python scripts that specify hardware components such as CPU type, memory system, cache configuration---and the workload that the simulated system will run. gem5 is compiled for its target instruction set architecture (ISA), such as x86, and executed using the aforementioned Python configuration scripts.

\subsection{Simulated hardware}
We profile the three major CPU types of gem5: AtomicSimpleCPU (AS), TimingSimpleCPU (TS) and Out-of-order (O3) CPU \cite{gem5_documentation}. Each Simple CPU class inherits from its from the parent class BaseSimpleCPU, which contains basic functionalities that each \emph{simple} CPU uses. In turn the BaseSimpleCPU inherits from its parent class: the BaseCPU. Due to the polymorphism of the C++ object-oriented code, certain CPU functions appear as belonging to these parent classes instead of the specific CPU.
\subsubsection{Atomic Simple CPU}
The \textbf{AS CPU} uses atomic memory accesses, \textit{i.e.} instructions are implemented as chains of function calls from a CPU into the memory, without decoupling the CPU and the memory subsystem. The AS CPU executes instructions in-order and without delay. Each Atomic packet used by the CPU will be owned by the requester, and a responder such as the memory system will modify the packet and embed the requested data. Thus, a request packet in the AS CPU will be completed through the returns of functions. The AS CPU lacks cycle accuracy, unlike CPUs using \emph{timed memory accesses} that decouple CPU from memory and model delays of the memory hierarchy. The AS CPU's straightforward execution scheme makes examining the execution flow simple since events are not delayed and subsequently enqueued, but instead directly executed via functions.
\subsubsection{Timing Simple CPU}
\label{desc:TSCPU}
The \textbf{TS CPU} models \emph{timing memory accesses}, \textit{i.e.} we decouple CPUs and memories by using memory subsystem events and wait for memory responses before proceeding. This is in constrast to the AS CPU, which would proceed immediately upon being able to. The TS CPU's timing is thus more realistic by including the latency introduced by internal queues or resource contention of the memory subsystem. Interactions consist of two one-way messages---a request will eventually receive a successful response ({\tt ACK}) or a negative acknowledgment ({\tt NACK}) if the request was not fulfilled.

The TS CPU sends a \emph{timing request} to the appropriate port's {\tt receive Timing} ({\tt recvTiming}) function. At the port, the request is either completed, or a NACK is sent back to the sender, which should then wait for a signal before trying again. Factors such as contention---multiple requests competing for the same resource---can cause packets to fail before reaching their destination, resulting in a NACK. If the sender needs a response, the responder may either modify the packet and return it, or delete the packet and generate a new one. The latter is necessary if the recipient is another target device than the requester. Memory devices who wish to reference the packet after it has been delivered must make a copy of it as it may be deleted and therefore invalid.

The TS CPU---like the AS CPU---executes instruction sequentially and therefore its stages are also: \emph{fetch instruction}, \emph{pre-execute} (decode), \emph{execute}, and \emph{post-execute}.

\subsubsection{O3 CPU}
The \textbf{O3 CPU} is gem5's implementation of a CPU using out-of-order execution and hence its most complex CPU model. It implements Tomasulo's algorithm \cite{tomasulo} to decouple CPU architectural states and internal execution states.

The O3 CPU uses \emph{timing memory accesses}---just like the TimingSimpleCPU (TS CPU); event-driven simulation where we decouple components such as the memory subsystem from each other. A simple CPU model of gem5 such as the AtomicSimpleCPU (AS CPU) or the TS CPU have four distinct execution stages: instruction fetch, pre-execute/instruction decoding, execute, commit. We briefly summarise each O3 CPU stage, however, more detail of each stage is provided alongside the profiling results in Section \ref{sec:O3_CPU_breakdown}.
\begin{itemize}
\item \textbf{Fetch}: In the \emph{Fetch} stage, the CPU selects a thread based on the provided policy. Furthermore, this is where a \emph{dynamic instruction} is constructed and where branch prediction is handled.
\item \textbf{Decode}: The \emph{Decoder} will take into account early resolution unconditional branches.
\item \textbf{Rename}: Instructions are assigned to physical registers, and we resolve register dependencies via renaming, which will map an architectural register to a physical one---preventing overwriting data and removing false dependencies\footnote{False dependencies such as \textbf{W}rite \textbf{A}fter \textbf{R}ead and \textbf{W}rite \textbf{A}fter \textbf{W}rite \emph{seemingly} introduce dependencies owing to their reuse of registers, however, only a true dependency such as \textbf{R}ead \textbf{A}fter \textbf{W}rite must be preserved. The use of physical registers allows for more flexible out-of-order instruction scheduling since instructions may be issued after their \emph{true} dependencies have been resolved.}.
\item \textbf{IEW}: The AS CPU and TS CPU have an \emph{execute} stage following its decoding. The O3 CPU instead has its IEW stage, which performs the out-of-order execution. IEW is a pipelined process with three stages: \textbf{I}ssue an instruction, \emph{i.e.} dispatching an enqueued instruction to start executing; \textbf{E}xecute, running ready instructions; \textbf{W}riteback, which stores instruction results before being committed.
\item \textbf{Commit}: The \emph{Commit} stage will make the instructions' actions externally visible, appearing as if they were executed in-order; ensuring program order regardless of execution order. It is also here any faults and branch misprediction are handled.
\end{itemize}

\subsubsection{\emph{Ruby}}
The memory subsystem handles memory management and transfers data packets throughout the simulated system. We use gem5's own \emph{\textbf{Ruby}} memory system. \emph{Ruby} allows for configuring cache coherence protocol implementations\footnote{The policy which determines the method of how to maintain cache coherence, i.e. uniformity of shared resources among multiple local caches.}, memory controllers, and other memory management aspects. In a \textbf{MESI} cache coherence protocol, which was used for this project, L1 cachelines can be in one of four states: \textbf{M}odified (\emph{dirty} cache line---inconsistent with memory; write-back to memory required), \textbf{E}xclusive (\emph{clean} and in a single cache), \textbf{S}hared (\emph{clean} and in multiple caches), or \textbf{I}nvalid (unusable) \cite{quantitative}. These states change upon either a Read/Write request from the cache line's owner, or a coherence request from another cache.

Coherence protocols in gem5 influences runtime performance, as its implementation may add significant latency during cache coherence operations. It is paramount to examine \emph{Ruby} since, as demonstrated later in Sections \ref{sec:AS_CPU_breakdown} and \ref{sec:TS_CPU_breakdown} that it constitutes a major portion of the total execution time of both the AS and TS CPUs. It is an open question what effect \emph{Ruby} has on gem5's simulation time---we address this for each CPU model.

For \emph{timing requests} of the TS CPU and O3 CPU, \emph{Ruby} employs its sequencer, which converts the CPU-side request into the corresponding \emph{Ruby}-side operation. This is event-driven since the sequencer will schedule asynchronous memory operations to emulate realistic timing and system behaviour.
\subsubsection{Garnet}
Garnet2.0 is the on-chip network model used in gem5. It is built upon the original Garnet \cite{garnet}, and it uses \emph{Ruby} for the topology and routing \cite{gem5_documentation}. Garnet simulates how packets move between memory subsystem components, \textit{e.g.} caches, CPUs or memory controllers in a network-on-chip (NoC)---an integrated communication system designed for multi-core processors. NoC replaces shared buses with a network of links and routers. Links connect the routers together, enabling \emph{Ruby} packets to travel between components. To enter Garnet, \emph{Ruby} packets pass through the network interface (NI), which converts them into network packets.

Garnet is not used in the AS CPU since it does not simulate timed interactions; its requests for \emph{Ruby} are completed once the function returns. The TS CPU and O3 CPU use timing accesses and therefore Garnet will simulate a packet traversing through the memory system.
\subsubsection{Request handling in \emph{Ruby} and Garnet}
We provide a brief description of how a timed request to \emph{Ruby} is handled. This is also illustrated in Figure \ref{fig:TS_anatomy}. The request will arrive at the appropriate {\tt receive} ({\tt recv}) function through \emph{Ruby}'s memory response port. It will be converted from a gem5 packet into a \emph{Ruby} Request Object, which is then sent to the interface of the current core's \emph{Ruby} \textbf{Sequencer}\footnote{This is assuming that the request is a normal memory request and not a Programmed I/O (PIO) request. PIO requests are routed to the correct PIO handler.}. The Sequencer will forward the request to the coherent cache hierarchy by enqueueing it in the \emph{mandatory queue} of the appropriate L1 cache controller. The L1 cache will eventually dequeue the request and perform a look-up---adjust its own coherence state transitions. 

If the requested cache block is found in the cache (hit), we immediately schedule a response; otherwise the L1 cache controller will push the request down the memory hierarchy or to a \emph{memory controller}. Requests are sent through \emph{message buffers}, which act as both an interface and a timing buffer between \emph{Ruby} components. The \emph{message buffer} stores the request until it is processed by the next component of the memory hierarchy. The request will need to traverse the NoC, and travel through Garnet routers.

Once the L1 cache has obtained the requested data, it will notify the Sequencer via a {\tt Callback}, of {\tt read} or {\tt write} type, depending on the request. The Sequencer will then call upon the \emph{Ruby} port with a {\tt hit callback}, which will deliver the memory response back to the requester. Specifically, a response will be scheduled in the Packet Queue and later sent to the appropriate {\tt recv} port. If the request was for an instruction, it is sent to the receive port of the instruction cache (I-cache). If the request was made by the data cache (D-cache), the response is sent to the D-cache's receive port.

\subsection{Simulated software}
\subsubsection{ISA}
gem5 simulates assembly-level instructions and supports a number of instruction set architectures (ISAs), which are abstract models describing the interactions between software and the CPU. More precisely, how the software controls the CPU. We used gem5's x86 ISA implementation---commonplace in commercial CPUs developed by Intel and AMD.
\subsubsection{gem5 instructions}
gem5 has different types of instructions. \emph{Static instructions}, which we denote as {\tt S-inst}, are unique instructions in the binary, and contain fixed properties, such as its op class, flags indicating instruction type, architectural register requirements and execution behaviour. During execution, an {\tt S-inst} is used to initially decode a binary instruction from the simulated program.

\emph{Dynamic instructions}, denoted as {\tt D-inst}, represent the actual execution of {\tt S-inst} and are necessary for more detailed models since they contain runtime properties, such as its program counter (PC), predicted next PC, which thread the instruction belongs to and the instruction's result. For example, each loop iteration will correspond to a new {\tt D-inst}, even though it is mapped to the same {\tt S-inst}. Furthermore, {\tt D-inst} provide the {\tt ExecContext} interface which allows instructions to interact with the CPU state: read or modify registers, access memory and update PC. Hence during execution, {\tt D-inst} uses the template of {\tt S-inst} to perform the execution in its actual dynamic context.

Finally, we collectively denote \emph{micro-operations} and \emph{macro-operations} as {\tt M-inst}. \emph{Macro-operations} are complex x86 instructions that are decomposed into simpler instructions called \emph{micro-operations} during execution. 
\subsubsection{System calls}
\begin{figure}[H]
    \centering
        \resizebox{0.37\linewidth}{!}{
    \begin{tikzpicture}[
        box/.style={
            draw,
            minimum width=5cm,
            minimum height=1cm,
            text centered,
            thick,
            rectangle,
            rounded corners
        },
        arrow/.style={
            draw,
            -{Stealth},
            thick
        }
    ]
        \node[
            box,
            fill=gray!20,
            minimum height=10cm,
            minimum width=7.5cm,
            label=above:{Host Machine}
        ] (host) at (0, -2.5) {};

        \node[
            box,
            fill=red!20,
            minimum height=2.5cm,
            minimum width=4.5cm,
            anchor=north
        ] (host_linuxapi) at (0, 1.66) {
            \vbox{
                \hbox{\textbf{Linux API}} 
                \vspace{1mm}
                \hbox{\begin{tikzpicture}
                    \node[
                        box,
                        fill=green!20,
                        minimum height=1.2cm,
                        minimum width=4cm
                    ] (host_kernel) {
                        \vbox{
                            \hbox{\textbf{Kernel}}
                            \hbox{\includegraphics[width=0.7cm]{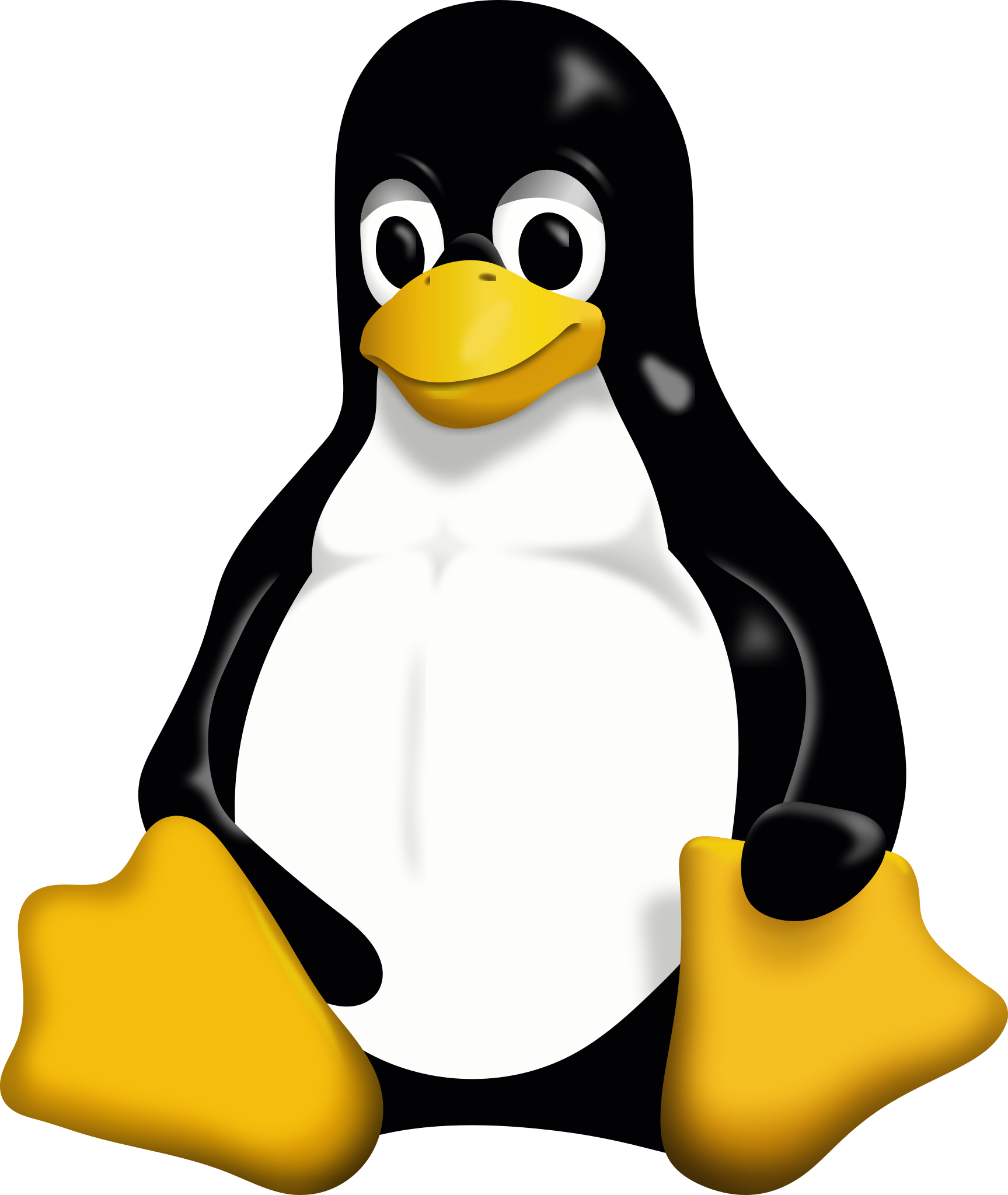}}
                        }
                    };
                \end{tikzpicture}}
            }
        };

        \node[
            box,
            fill=blue!20,
            minimum height=5cm,
            minimum width=6cm,
            anchor=north
        ] (gem5) at (0, -1.6) {
            \vbox{
                \hbox{\textbf{gem5}} 
                \vspace{5mm}
                \hbox{\begin{tikzpicture}
                    \node[
                        box,
                        fill=yellow!20,
                        minimum height=1.5cm,
                        minimum width=4cm
                    ] (application) at (0, 2.2) {
                        \vbox{
                            \hbox{\textbf{Application}}
                        }
                    };

                    \node[
                        box,
                        fill=red!20,
                        minimum height=2.5cm,
                        minimum width=4.5cm
                    ] (gem5_linuxapi) at (0, 0) {
                        \vbox{
                            \hbox{\textbf{Linux API}} 
                            \vspace{1mm}
                            \hbox{\begin{tikzpicture}
                                \node[
                                    box,
                                    fill=green!20,
                                    minimum height=1cm,
                                    minimum width=4cm
                                ] (gem5_kernel) at (0, -0.8) {
                                    \vbox{
                                        \hbox{\textbf{Kernel}}
                                        \hbox{\includegraphics[width=0.7cm]{./tux.png}}
                                    }
                                };
                            \end{tikzpicture}}
                        }
                    };
                \end{tikzpicture}}
            }
        };

\path[
    draw=black,
    ultra thick,
    -{Latex[length=3mm, width=3mm]}
] (gem5.north) to (host_linuxapi.south);

\path[
    draw=red!60,
    very thick,
    -{Latex[length=3mm, width=3mm]}
] ([xshift=-2mm]host_linuxapi.south) to ([xshift=-2mm]gem5.north);

\draw[
    draw=black,
    very thick,
    -{Latex[length=3mm, width=3mm]}
] (0,-3.9) -- (0,-4.8);

\draw[
    draw=red!60,
    very thick,
    -{Latex[length=3mm, width=3mm]}
] (0.3,-4.8) -- (0.3,-3.9);

    \end{tikzpicture}
    }
    \caption{Illustration showing how gem5 running full system mode interacts with the mounted kernel and the host machine's kernel.}
    \label{fig:kernelsFS}
\end{figure}

\begin{figure}[H]
    \centering
    \resizebox{0.37\linewidth}{!}{
    \begin{tikzpicture}[
        box/.style={
            draw,
            minimum width=5cm,
            minimum height=1cm,
            text centered,
            thick,
            rectangle,
            rounded corners
        },
        arrow/.style={
            draw,
            -{Stealth},
            thick
        }
    ]
        \node[
            box,
            fill=gray!20,
            minimum height=10cm,
            minimum width=7.5cm,
            label=above:{Host Machine}
        ] (host) at (0, -2.5) {};

        \node[
            box,
            fill=red!20,
            minimum height=2.5cm,
            minimum width=4.5cm,
            anchor=north
        ] (host_linuxapi) at (0, 1.66) {
            \vbox{
                \hbox{\textbf{Linux API}} 
                \vspace{1mm}
                \hbox{\begin{tikzpicture}
                    \node[
                        box,
                        fill=green!20,
                        minimum height=1.2cm,
                        minimum width=4cm
                    ] (host_kernel) {
                        \vbox{
                            \hbox{\textbf{Kernel}}
                            \hbox{\includegraphics[width=0.7cm]{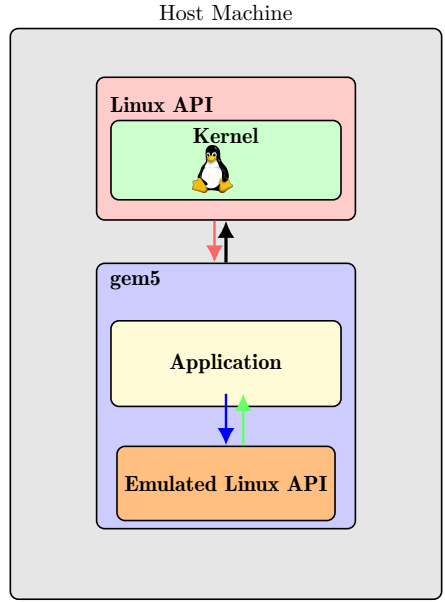}}
                        }
                    };
                \end{tikzpicture}}
            }
        };

        \node[
            box,
            fill=blue!20,
            minimum height=2.5cm,
            minimum width=4.5cm,
            anchor=north
        ] (gem5) at (0, -1.6) {
            \vbox{
                \hbox{\textbf{gem5}} 
                \vspace{5mm}
                \hbox{\begin{tikzpicture}
                    \node[
                        box,
                        fill=yellow!20,
                        minimum height=1.5cm,
                        minimum width=4cm
                    ] (application) at (0, 2.2) {
                        \vbox{
                            \hbox{\textbf{Application}}
                        }
                    };

                    \node[
                        box,
                        fill=orange!50,
                        minimum height=1.3cm,
                        minimum width=2.5cm
                    ] (gem5_linuxapi) at (0, 0) {
                        \vbox{
                            \hbox{\textbf{Emulated Linux API}} 
                        }
                    };
                \end{tikzpicture}}
            }
        };

\path[
    draw=black,
    ultra thick,
    -{Latex[length=3mm, width=3mm]}
] (gem5.north) to (host_linuxapi.south);

\path[
    draw=red!60,
    very thick,
    -{Latex[length=3mm, width=3mm]}
] ([xshift=-2mm]host_linuxapi.south) to ([xshift=-2mm]gem5.north);

\draw[
    draw=blue,
    very thick,
    -{Latex[length=3mm, width=3mm]}
] (0,-3.9) -- (0,-4.8);

\draw[
    draw=green!60,
    very thick,
    -{Latex[length=3mm, width=3mm]}
] (0.3,-4.8) -- (0.3,-3.9);

    \end{tikzpicture}

    }
    \caption{Illustration of how gem5 running syscall emulation mode interacts with the Linux kernel of the host machine.}
    \label{fig:kernelsSE}
\end{figure}

Computer programs request services to its host operating system via system calls (syscalls). This can involve managing processes or accessing hardware components such as the disk or a peripheral device. It is in general provided as an application programming interface (API). In operating systems such as the Linux kernel, its API allows a program running in userspace (limited privileges) to make syscalls to the kernel \cite{linux_syscalls_man_page}, thus accessing system resources and services of the kernelspace (higher privileges).

gem5 can execute in two modes: full system (FS) and syscall emulation (SE). In FS, gem5 will emulate the entire system; including running a complete kernel. Whereas in SE, userspace is the focus and syscalls are emulated. The tradeoff is that SE is simpler to configure and presumably faster, whereas FS is more accurate to true execution but requires further setup, including mounting a Linux kernel. As gem5 itself is a program running on a host machine, it will also make syscalls to its host's kernel. We illustrate this distinction between the mounted kernel and the host's kernel in Figure \ref{fig:kernelsFS}, showing FS mode. In contrast, SE mode will solely simulate userspace code, and thus uses a simple emulated Linux API, as illustrated in Figure \ref{fig:kernelsSE}. We solely use FS mode in this project since the simulation is more representative to execution on real hardware.

\section{Prior work on gem5 profiling}
\label{sec:prior}
We believe that characterising gem5's runtime behaviour can be valuable to the gem5 community---this may involve identifying its bottlenecks, examining configurations' performance, how gem5's internal components interact and how the simulator interacts with its host system. Such insights may guide the simulator's development as well as help beginners with an eased learning curve. Yet, little work related to this topic has been published. Umeike \textit{et al.} \cite{umeike2023profiling} profiled gem5 to assess its bottlenecks in terms of the host machine's hardware. They used software such as Intel VTune \cite{vtune} to perform a top-down microarchitectural analysis to illustrate that gem5's execution time was sensitive to smaller L1 cache sizes along with smaller virtual memory page sizes. Furthermore, they conclude that using an M1 CPU had the potential of reducing simulation time significantly when compared to a high-end machine using an Intel Xeon CPU. Their paper claims that there is no \emph{killer function} inside gem5 for its CPU types, based on their observation that simulation time will simply increasing with the number of functions.

The insights of their paper are valuable in terms of what host machine hardware is optimal for gem5, however, the question of what is going on internally in gem5 remains largely unanswered. This report therefore aims to address this; we provide insights into how gem5 configurations impact performance, its software bottlenecks and how each CPU type interacts with the \emph{Ruby} memory subsystem.

\begin{figure}[p]
  \includegraphics[width=0.9\textwidth, height=0.51\textheight]{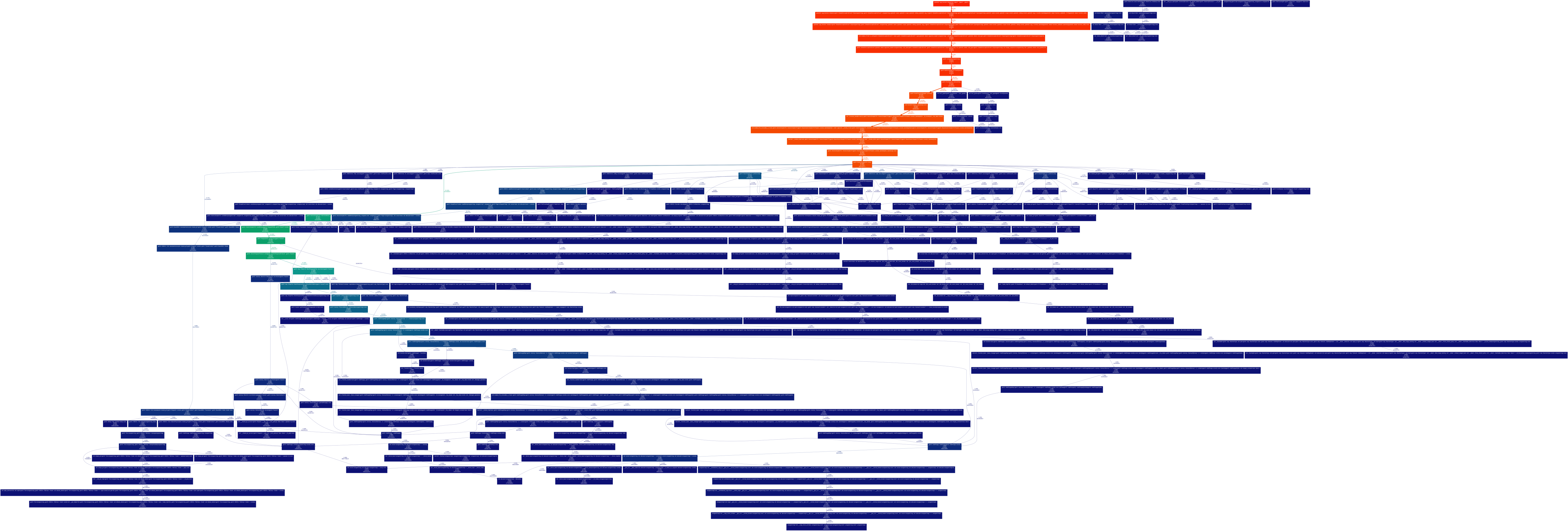}
  \caption{\label{gprof_60min_callgraph} Example of a gprof callgraph, generated by gprof2dot on gprof data obtained from running gem5 for 60 minutes on the Blackscholes benchmark of Parsec-3.0, using the \emph{Ruby} memory subsystem, 3GB of memory and 1 core.}
\end{figure}

As of 2025, gem5 officially supports building options to enable profiling with gprof. Using gprof2dot \cite{gprof2dot}, a library used for parsing profiling data from tools such as gprof, we can generate a graphical image of all function calls in gem5.
Figure \ref{gprof_60min_callgraph} shows such an image after running {\tt blackscholes} from PARSEC-3.0 for $60$ minutes.

One key shortcoming of gprof is that the gprof structure is flattened such that for a callee function with multiple callers, gprof's graph will have edges to each of the callers, making the picture unnecessarily complex. It is therefore not easy to interpret gprof output, especially when the user wants to focus on a specific function call chain amid many others.
To preserve the hierarchical callee/caller structure of each function call and to be able to flexibly examine the profiling data, we develop our own profiling tool, detailed in Section \ref{sec:profiler}.
Our profiler supports user-defined configuration files to parse the profiling results of interests, ignoring the uninteresting others, and extracting data to generate stacked bar charts demonstrating the execution time breakdown for the relevant gem5 components.

\begin{figure}[h]
  \includegraphics[scale=0.1]{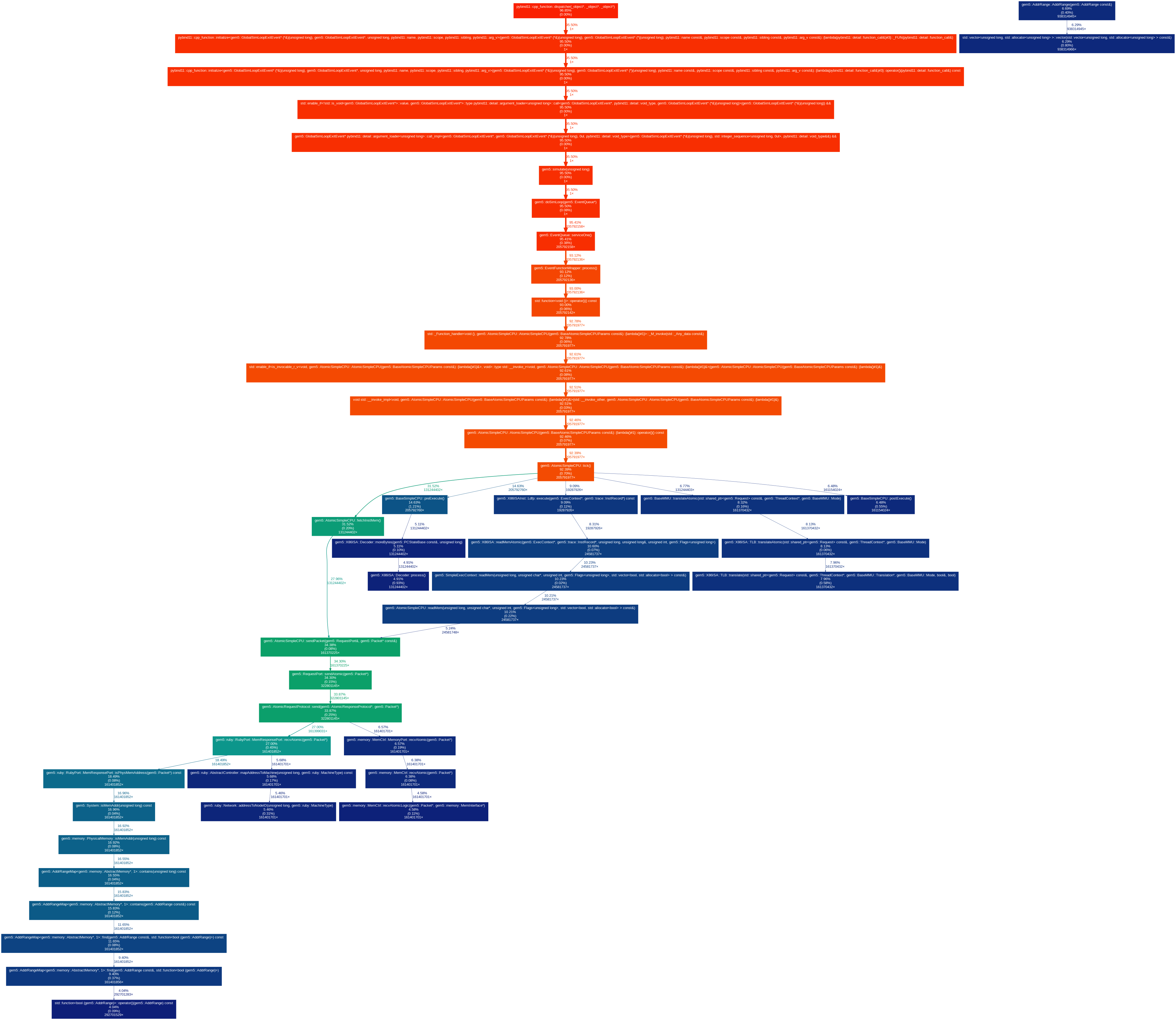}
  \caption{\label{gprof_60min_callgraph_4p} Example of a gprof callgraph, generated by gprof2dot on gprof data obtained from running gem5 for 10 minutes on the Blackscholes benchmark of Parsec-3.0, using the \emph{Ruby} memory subsystem, 3GB of memory and 1 core. \textbf{Instructions account for less than 4\% of the total runtime have been omitted}.}
\end{figure}

\section{Purpose and aims of this report}
\label{sec:purpose-goals}
The prior profiling work discussed in Section \ref{sec:prior} does not address the internal mechanisms of gem5 and instead solely focuses on how to accomodate hardware for the current gem5 software. Umeike \textit{et al.} \cite{umeike2023profiling} claim that there is no \emph{killer function} inside gem5, which does not consider gem5's modular structure---a more appropriate topic to investigate instead may be whether there exists a \emph{killer component} in gem5. The gprof data is unsorted and shows a single complete callgraph---any meaningful results are therefore difficult to decipher since all gem5 layers are presented together. This report aims to specifically address these gaps: we reveal the internal behaviour of gem5's simulated hardware and software in a modular manner mirroring how gem5 itself, as a software project, is designed. The profiling results are presented to explain components' interactions and the execution flow along with demonstrating how execution time is partitioned at each layer of gem5. Particularly, this report aims to answer the following questions:
\begin{itemize}
\item What impact do configuration choices have on gem5's execution behaviour and runtime?
\item How does each analysed CPU interact with the broader system and what are their performance bottlenecks?
\end{itemize}
We run gem5 on a selection of popular benchmarks: SPEC2017 \cite{spec2017}, PARSEC-3.0 \cite{parsec3}, and GAPBS \cite{GAPBS}.

Our work establishes a clear performance baseline against which a gem5 component designer can \emph{verify whether an integrated hardware module functions as intended}. For example, in a project aiming to improve the memory management unit (MMU) design, our methodology can deliver fine-grained results to assess whether the new MMU performs as expected or if it introduces unintended effects that trigger anomalous hardware behaviour. Furthermore, our methodology \emph{reveals the application’s behaviour}—its hardware usage and instruction set characteristics—providing crucial insights for deeper analysis, such as uncovering inefficiencies in hardware utilization.

\chapter{Methodology}
\label{cha:method}
A machine with the following specification was used to obtain the results presented in Sections \ref{sec:AS_CPU_breakdown}, \ref{sec:TS_CPU_breakdown} and \ref{sec:O3_CPU_breakdown}: \specs. We present the profiling results of each CPU at each layer in with figures produced using data aggregated by the parser along with a description of the execution flow.
\section{Benchmarks and gem5 configuration}
There are three binary options when compiling gem5: \emph{fast} (optimisation flags enabled, debug symbols disabled; no function symbols; asserts are skipped), \emph{opt} (optimisation flags, debug symbols, and asserts are enabled), and \emph{debug} (optimisation flags disabled, debug symbols and asserts are enabled). It is recommended to use \emph{fast} for the average user since they may not need debugging information, and its inclusion slows down the simulation. For this project, we modified the compilation flags of the \emph{fast} binary to include debug symbols, in particular function names as to have eligible function callgraphs.

Our benchmarks are detailed in Table \ref{tab:benchmarks}. The gem5 configurations are defined in terms of CPU type, the number of cores, amount of memory and memory subsystem. Each benchmark has its own set of configurations, specificed in Table \ref{ta:cfg_core} for the number of cores and Table \ref{ta:cfg_mem} for the amount of memory. We solely use \emph{Ruby} for the memory subsystem with a {\tt MESI\_Two\_Level} configuration, which uses the MESI protocol for a two-level cache hierarchy, \textit{i.e.} ensuring consistency between its L1 caches and a shared L2 cache.

All experiments are conducted from the beginning of each benchmark’s region of interest (ROI). For PARSEC and GAPBS, the ROI is defined at the entry to the parallel phase following initialization. For SPEC, we apply the simpoint methodology \cite{simpoint}, selecting checkpoints that represent the most intensive phase of each program. 

\begin{table}[htbp!]
 \centering
\begin{tabular}{|c|p{10cm}|} 
 \hline
 \textbf{Benchmark suite} & \textbf{Applications} \\ 
 \hline
 \textbf{GAPBS}           & {\tt bc\_raw}, {\tt bfs\_raw}, {\tt cc\_raw}, {\tt cc\_sv\_raw}, {\tt pr\_raw}, {\tt pr\_spmv\_raw}, {\tt tc\_raw} \\ 
 \hline
 \textbf{PARSEC-3.0}       & {\tt blackscholes}, {\tt bodytrack}, {\tt dedup}, {\tt facesim}, {\tt ferret}, {\tt fluidanimate}, {\tt freqmine}, {\tt raytrace}, {\tt streamcluster}, {\tt swaptions}, {\tt vips} \\ 
 \hline
   \textbf{SPEC2017}         & {\tt bwaves}, {\tt cactuBSSN}, {\tt exchange2}, {\tt gcc}, {\tt imagick}, {\tt lbm}, {\tt leela}, {\tt mcf}, {\tt nab}, {\tt omnetpp}, {\tt perlbench}, {\tt x264}, {\tt xalacbmk} \\ 
 \hline

\end{tabular}
\caption{The Benchmark suites: GAPBS \cite{GAPBS}, PARSEC-3.0 \cite{parsec3}, and SPEC2017 \cite{spec2017}, and their applications used in this project.}
\label{tab:benchmarks}
\end{table}
\begin{table}[htbp!]
  \centering
\begin{tabular}{|c|| c|| c|| c|} 
 \hline
 \textbf{Benchmark suite} & \textbf{1 core} & \textbf{4 cores} & \textbf{16 cores} \\
 \hline
 GAPBS      & \cmark &  \cmark & \cmark \\ 
 \hline
 PARSEC-3.0 & \cmark &  \cmark  & \cmark \\
 \hline
 SPEC2017 & \cmark & \xmark & \xmark \\
 \hline
\end{tabular}
\caption{The benchmark suites and their configurations in terms of simulated cores.}
\label{ta:cfg_core}
\end{table}
\begin{table}[htbp!]
  \centering
\begin{tabular}{|c|| c|| c|| c|| c|| c|} 
 \hline
 \textbf{Benchmark suite} & \textbf{3GB} & \textbf{8GB} & \textbf{16GB}\\
 \hline
 GAPBS      & \cmark & \cmark & \xmark \\ 
 \hline
 PARSEC-3.0 & \cmark & \xmark & \cmark \\
 \hline
 SPEC2017 & \cmark & \xmark & \xmark   \\
 \hline
\end{tabular}
\caption{The benchmark suites and their configurations in terms of simulated memory.}
\label{ta:cfg_mem}
\end{table}
\section{A new profiler for profiling gem5}
\label{sec:profiler}
An in-house profiler is employed to profile gem5. It is provided the process identifier (PID) of the gem5 simulation, and will at a regular interval collect data in the form of a function callchain, ending at the currently running function\footnote{The sampling interval was set to $1000$ milliseconds on this project.}. Each sample is added to a JavaScript Object Notation (JSON) structure, which can be parsed as a Python dictionary (dict). These JSON files are later processed by an in-house parser, described in Section \ref{sec:parser} to produce the graphs shown in Sections \ref{sec:AS_CPU_breakdown}, \ref{sec:TS_CPU_breakdown} and \ref{sec:O3_CPU_breakdown}.

The profiler uses the {\tt perf\_event\_open} system call (syscall) of the Linux kernel to access performance counters and monitor kernel events. The syscall takes the following arguments: an {\tt attr}-struct\footnote{Defined in the Linux kernel as a {\tt perf\_event\_attr}.}, defining the event to monitor and the monitoring configuration, a process identifier (PID), which CPU to monitor, event group membership, and optional flags.

For each sample {\tt perf\_event\_open} collects, a function callchain is produced, \textit{i.e.} callstack of function calls. This structure may also be described as a \emph{tree} since software function calls often share the same root node. The callchain is traversed bottom-up and each callchain is stored in a nested JSON structure. It is essentially a reverse depth-first search, starting at the bottom and moving towards the root node; the instruction that boots the program.

If a function callchain has not been previously recorded, it is added to the JSON structure and given the initial sample count of $1$; if a callchain has been previously recorded, its sample count is incremented. Hence, each function call of the entire JSON object has a ``count'' attribute succeeding a potential nested JSON object containing its children\footnote{It does not have any children aside from the ``count'' attribute if it is a leaf node.}. Each function call's portion of the total samples is calculated like this: $100*\frac{sample\_count}{total\_samples}$\footnote{Where $total\_samples$ is equivalent to the root instruction's number of samples.}. A {\tt callstack.json} file with the data is generated once the profiler terminates. We produced one such file for each gem5 configuration, for each benchmark application. Each file is stored in a directory structure corresponding to the configuration and bench in the following order: \begin{enumerate}
\item Benchmark name.
\item Application name.
\item Number of cores.
\item CPU type.
\item Amount of memory.
\item If the \emph{Ruby} memory subsystem was used, the file is placed in a \emph{ruby} directory; otherwise it is placed one level higher.
\end{enumerate}
An example of this would be: {\tt parsec-3.0/blackscholes/1/AtomicSimpleCPU/3GB/ruby/} -- the path to this configuration's {\tt callstack.json} file. We run each execution (configuration and application combination) for $1$ hour with the profiler collecting samples throughout. This means that we cannot assess the effect of total execution time since all executions were done in $1$ hour; instead, we can only make observations on the \emph{relative time spent inside each component} compared across configurations. We imposed this time limit to accomodate the time constraints of the project. $458$ JSON files were used in total; generated from $458$ CPU hours, but were executed concurrently in limited batches. Considering that it is not unusual for a single execution to require more than $10$ hours to complete, the decision was made to set the limit to $1$ hour in order to collect the needed data in an achievable time. This additionally allows for concretely comparing the division of execution time for all executions across a comparable time frame.
Further, to ensure that our profiler targets only the gem5 process, we create a dedicated cgroup for each gem5 run.

\section{Parsing the profiling results}
\label{sec:parser}
The parser used was developed by us and is used to generate figures out of the parsed data. The parser will take a configuration file as an argument, which will specify which root function to parse from. The data tree is traversed until the matches of the provided root is found. Each child of each matching root will be included in the intermediate parsing result\footnote{The same function may be called at different locations in the callstack. Furthermore, as the case of gprof, the parser also has the ability to \emph{flatten} the call graph, \textit{i.e.} parse for leaf nodes instead of children of the root.}. The configuration specifies a whitelist and a blacklist respectively. The whitelist will associate a child with a specified category, as multiple function calls may be grouped together based on a shared purpose or general category.
The blacklist omits matching children from the parsed result, which is necessary because the profiler often yields trivial function names from Python---pybind11 C++ support used for gem5’s Python-based configuration.

To accurately categorise each child of the whitelist, the gem5 source code was studied in order to comprehend each function's role. Based on the source code, a number of configuration files were written for the parser to produce the bar plot figures in Sections \ref{sec:AS_CPU_breakdown}, \ref{sec:TS_CPU_breakdown} and \ref{sec:O3_CPU_breakdown}.
For each configuration executed on a given application, the figure label is given in the following format: for instance, {\tt 1AS3r} denotes: \textbf{1} \textbf{AS} core, \textbf{3} GB of memory and running \emph{\textbf{r}uby}.

\chapter{Results for Atomic Simple CPU}
\label{sec:AS_CPU_breakdown}
\begin{figure}[p]
  \centering
    \rotatebox{90}{
\includegraphics[width=1.4\textwidth]{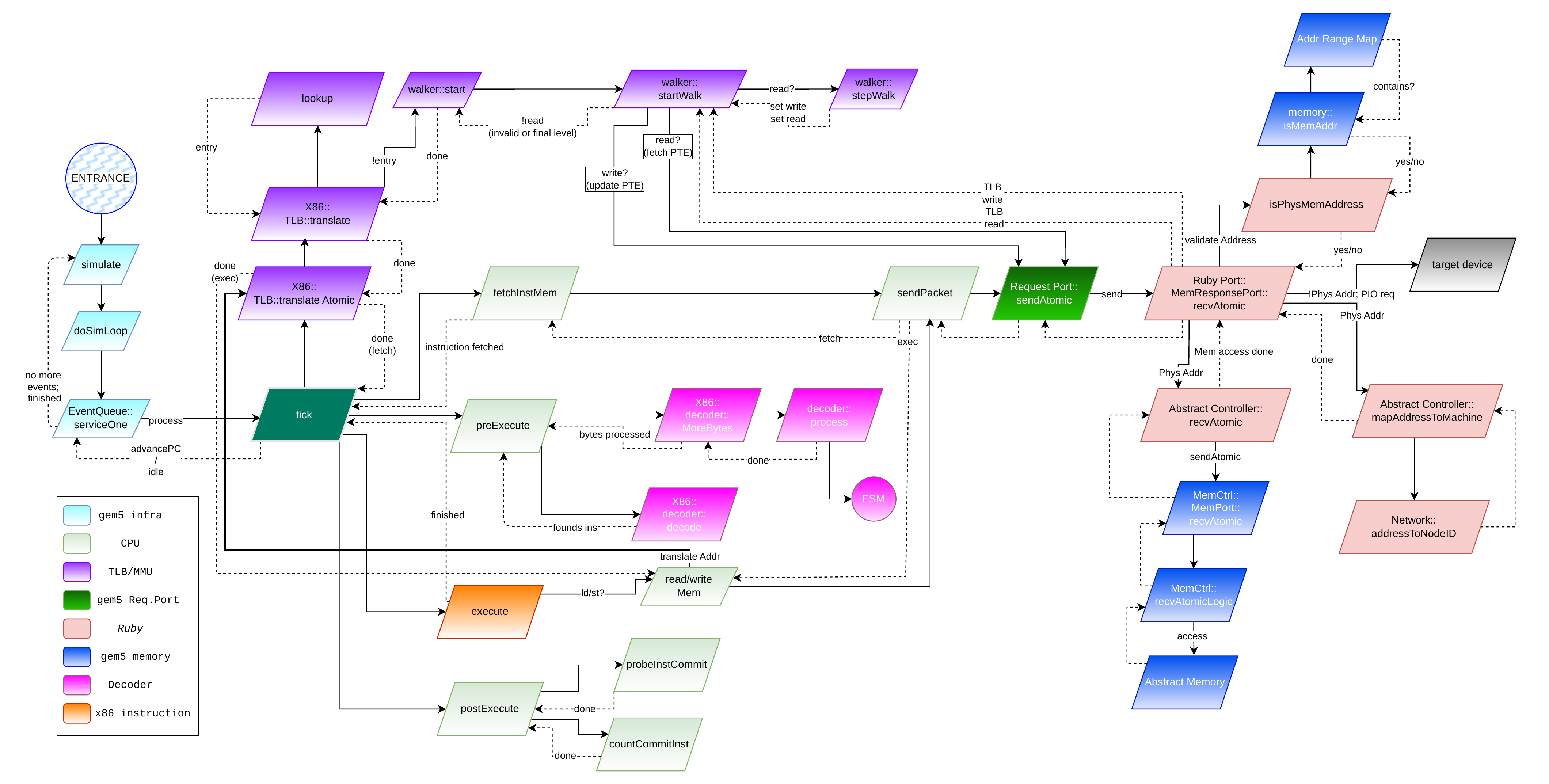}
}
\caption{\label{fig:AS_anatomy} Anatomy figure showing the function call interactions between the components driven by the AS CPU.}
\end{figure}

At the top-level of the AS CPU is the \emph{tick} function, which is called every CPU cycle. We present a function call anatomy of the AS CPU in Figure \ref{fig:AS_anatomy}, which contains each execution stage analysed in subsequent sections. CPU functions are denoted by green boxes; light blue denotes internal gem5 simulation management; purple denotes the MMU/TLB; dark blue denotes the gem5 memory parent class; light red denotes \emph{Ruby}; pink denotes the decoder and orange denotes x86 instruction.
We present the division of execution time inside \emph{tick} in \profilingRefs{AS-overall}---{\tt D-inst} denotes \emph{dynamic instructions}; {\tt S-inst} denotes \emph{static instructions}; {\tt Core-other} denotes miscellaneous operations of the CPU; {\tt advancePC} moves the \emph{program counter}; {\tt postExec} denotes post-execute stage; {\tt preExec} denotes the pre-execute stage; {\tt fetch} denotes the fetch instruction stage; {\tt gem5-infra} denotes internal gem5 instructions; {\tt Context} denotes instructions related to \emph{context switching}; {\tt Decoder} denotes the \emph{instruction decoder} and {\tt M-inst} denotes macro- and micro-instructions.

We observe that the overall execution time spends a significant portion on instruction handling, \textit{i.e.} operations related to instruction fetching ({\tt fetch}), decoding ({\tt preExec}) and post-execution ({\tt postExec}). {\tt fetch} is consistently the most prominent operation among them. Depending on the workload, instruction preperation constitutes either half of, or the majority of the total execution time\footnote{\emph{Instruction preperation} defined as the combined execution time of {\tt fetch} and {\tt preExec}.}.

Dynamic instruction execution tends to either take up an equal amount of execution time as {\tt fetch} or less than it, with one notable exception being {\tt lbm} of Spec2017, which is heavily dominated by {\tt D-inst}. Static instructions are generally much less prominent than their dynamic counterparts. The remaining execution time is dedicated to various operations, including the memory management unit (MMU) and internal operations of the CPU and gem5. Each portion of the execution time is examined more in detail inside its dedicated section. Due to gem5's object-oriented code structure, several higher layers will have small portions of pure C++ functions being executed.
\profilingFigs{AtomicSimpleCPU}{AS-overall}{\emph{tick} function's execution division}
\section{Instruction fetch}
\label{section:fetchins}
The Atomic CPU will send a fetch request through its request port, which will arrive at \emph{Ruby}'s memory response port, as illustrated in Figure \ref{fig:AS_anatomy}---\profilingRefs{AS-fetch} illustrate the division of execution time upon sending a \emph{fetch instruction} request, and that from the point of view of the AS CPU's request port, fetching an instruction from memory to the CPU is dominated by the \emph{Ruby} memory subsystem. As we show in Figure \ref{fig:AS_anatomy}, \emph{Ruby} is used when both fetching an instruction and to execute a {\tt Load} or {\tt Store}. Both instruction fetching and execution spend the vast majority of their respective execution time inside \emph{Ruby}; therefore, we solely examine \emph{Ruby} in Section \ref{sec:AS_res_ruby}.
\profilingFigs{AtomicSimpleCPU}{AS-fetch}{Division of execution {\tt sendAtomic}}

\section{Pre-execute}
\label{section:preexecute}
Once the CPU has fetched an instruction, it must be decoded in the Pre-execute stage. The majority of the execution time is spent inside {\tt moreBytes}, fetching data that it is then incrementally decoded. Once an instruction is ready, {\tt decode} is invoked to retrieve the decoded instruction. \profilingRefs{AS-preExec-decoder} show the division of execution time inside the top-level {\tt preExecute} function. We observe that {\tt decode} takes up more than a third of the execution time and that the {\tt moreBytes} itself makes up the vast majority of the remainder.
\profilingFigs{AtomicSimpleCPU}{AS-preExec-decoder}{Division of execution for the decoder}
\paragraph{\textbf{\emph{Impact of instruction set diversity and core scaling on decoding performance}}:}
Increasing memory correlates reducing the {\tt preExec} portion of the overall runtime division summary in Figures \ref{fig:AS-overall_gapbs} and \ref{fig:AS-overall_parsec3}. In Figure \ref{fig:AS-preExec-decoder_gapbs} we can see that increasing memory correlates to relatively less time spent on {\tt moreBytes}. This may be due to {\tt decode} benefitting from the improved locality of increased memory across all applications.

The impact of increasing the number of cores is inconclusive. For example in PARSEC-3.0, certain applications such as {\tt blackscholes} and {\tt raytrace} experience relatively longer execution time inside {\tt moreBytes} when increasing the number of cores, whereas {\tt dedup}'s relative time is decreased. This may be due to that it uses a more diverse set of instructions that benefits from parallelisation. Whereas both {\tt blackscholes} and {\tt raytrace} are comparatively homogeneous, as seen in Figure \ref{fig:AS-inst_parsec3} -- they are dominated by {\tt Ldfp}.

In GAPBS, the majority of applications are dominated by {\tt Ldbig} and {\tt Ldfp} (see Figure \ref{fig:AS-inst_gapbs}) and we observe no consistent effect by increasing the number of cores. {\tt cc\_raw} has a relatively prominent portion of {\tt Store} ({\tt St}). Counter to the hypothesis that diverse instruction sets improve parallelism, {\tt cc\_raw} does not spend noticeably less relative time inside {\tt moreBytes} with more cores. Nevertheless, it should be noted that we only consider {\tt cc\_raw} \emph{``diverse''} in the context of GAPBS and it is much less diverse than certain applications of PARSEC, such as {\tt dedup}.

{\tt bc\_raw} of GAPBS is the most diverse application of the benchmark suite, and it \emph{does} spend less time on {\tt moreBytes} when increasing from $1$ to $4$ cores, but this trend does not continue for $16$ cores. {\tt tc\_raw}, while less diverse than {\tt bc\_raw}, has a slightly more consistent runtime trend, but again it is moreso impacted significantly by increasing memory than cores.

\begin{figure}[H]
    \centering
    \includegraphics[width=\textwidth]{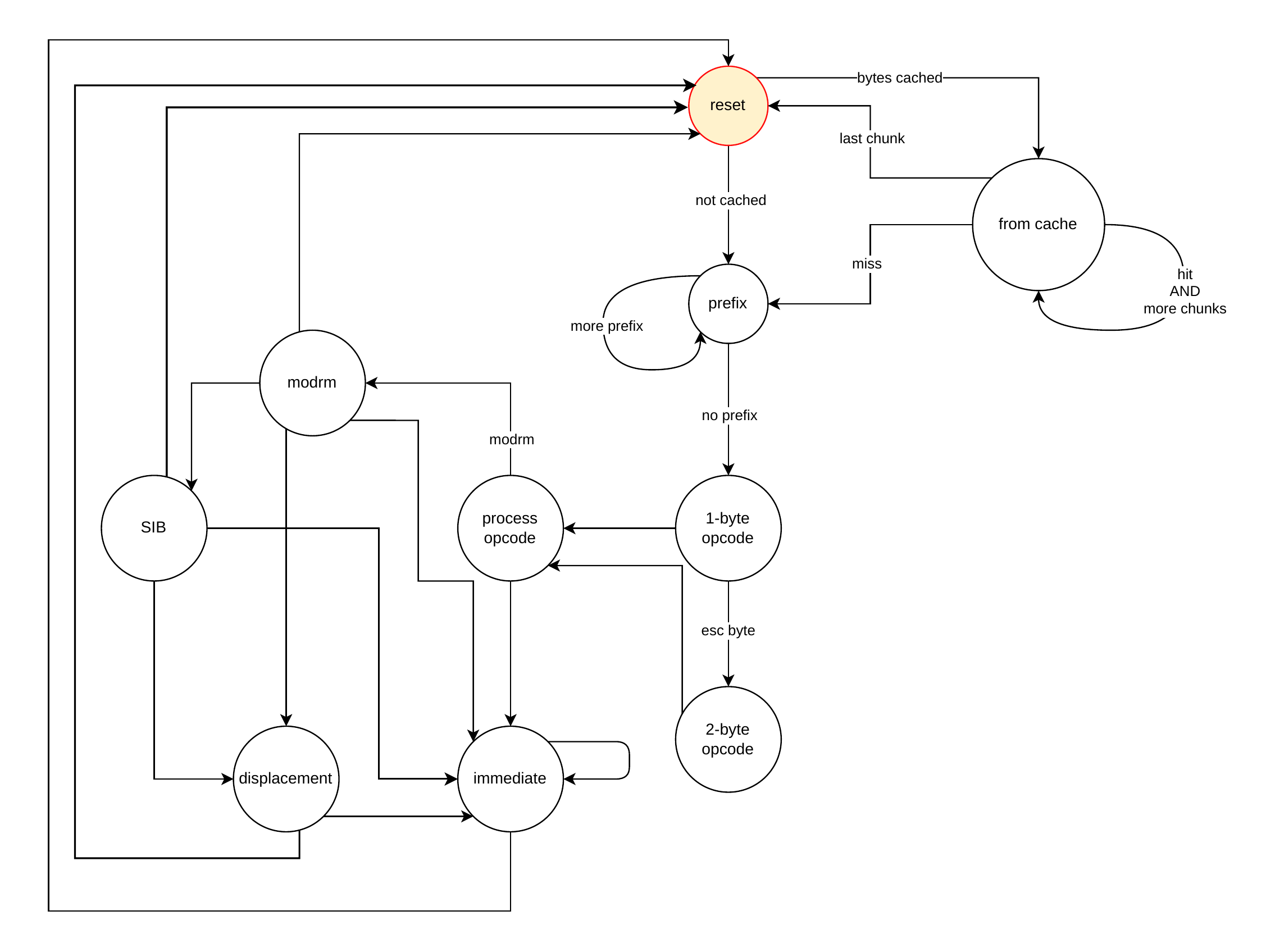}
    \caption{\label{fig:processFSM} Simplified FSM of the {\tt process} function.}
  \end{figure}

{\tt fluidanimate} of PARSEC and {\tt pr\_raw} of GAPBS have a similar instruction set dominated by {\tt Ldfp} and {\tt LdBig}. {\tt fluidanimate} spends slightly relatively less time on decoding going from $1$ to $4$ cores, but going from $4$ to $16$ cores makes it slightly slower. {\tt pr\_raw} becomes marginally slower going from $1$ to $4$ cores, and then at $16$ cores is essentially the result as at $1$ core. This suggests that instruction set alone is not a sufficient indicator of the configuration's impact on the execution time of {\tt decode} -- it is likely a combination of instruction sets and internal control flow of the application.
\paragraph{\emph{\textbf{Process FSM bottleneck}}:}
{\tt moreBytes} will call upon the decoder's {\tt process} function, which utilises a finite state machine (FSM) to manage its task. We examine the FSM used to implement {\tt process} of the decoder. Figure \ref{fig:processFSM} is a simplified illustration of the FSM, excluding certain states that were not prominent in the profiling data\footnote{Excluded states are chiefly related to Vector Extension (VEX) -- used for Single Instruction, Multiple Data (SIMD) instructions. We excluded them to retain a relatively simple illustration and since they were not prominent throughout the profiling data.}. We briefly explain each state below.
\begin{itemize}
\item {\tt \textbf{reset}} -- initial state; determines if bytes are cached.
\item {\tt \textbf{prefix}} -- instruction prefixes; loops until each prefix is processed, then transitions to opcode decoding states.
\item {\tt \textbf{from cache}} -- validates each cached instruction chunk.
\item {\tt \textbf{1-byte opcode}} -- decodes 1-byte of opcodes; transitions to the 2-byte state if an escape byte is found\footnote{An \emph{escape byte} is a special indicator that additional opcode bytes will follow.}, otherwise it transitions to the process state.
\item {\tt \textbf{2-byte opcode}} -- decodes 2-byte opcodes; transitions to the process state once finished.
\item {\tt \textbf{process opcode}} -- determines the next step based on the encountered data.
\item {\tt \textbf{modrm}} -- decodes the ModR/M byte for determining adressing mode, register operands or memory references. It also determines if other steps are needed before completing by transitioning to reset.
\item {\tt \textbf{SIB}} -- handles the Scale-Index-Base (SIB) byte for complex memory addressing, and appears after the ModR/M byte. Once finished, it determines the next step.
\item {\tt \textbf{displacement}} -- collect memory offset bytes for instructions using relative addressing; decides if an immediate is required.
\item {\tt \textbf{immediate}} -- collect immediate operand bytes of an instruction such as a constant\footnote{Also known as \emph{immediate values}.}; once complete, transition to reset.
\end{itemize}
{\tt Done} is invoked upon finishing processing a byte and then transitioning back to {\tt Reset}. {\tt getNextByte} is invoked at each state transition and is the backend driving the FSM -- if for instance {\tt immediate} processes three immediate bytes for a give instruction, then {\tt getNextBytes} is invoked three times at that state.
We observe that {\tt Reset} is the most time-consuming state of the FSM in \profilingRefs{AS-DP}. It accesses the cache to see whether the current bytes are present or not---this is likely its pitfall performance-wise. It is therefore intuitive that {\tt Reset} is the most time-consuming, since regardless of which instruction is decoded, each will go through the setup involving the cache check. 
\profilingFigs{AtomicSimpleCPU}{AS-DP}{Execution division of {\tt decode}'s FSM}
Taking all benchmarks into account, {\tt 1-byte opcode}, {\tt prefix} and {\tt modrm} tend to share second place of most time-consuming state. For PARSEC-3.0, {\tt 2-byte opcode} is the additional third contender of being second as it has much more prominence than in GAPBS. They all depend heavily upon gem5's {\tt BitfieldType}---its interface for C++ bit operations. Their order in the time-consumption hierarchy makes sense since they each are critical across the instructions and 2-byte opcodes have a relatively larger presence in PARSEC-3.0. Less prominent states such as {\tt Immediate} are indeed not time-consuming owing to their simple operations, such as typically reading a byte and determining the next path in the FSM. Increasing the memory from 3GB to 16GB for PARSEC-3.0 and from 3GB to 8GB on GAPBS, slightly decreases the relative runtime of {\tt Reset} and {\tt ModRM}.

\profilingRefs{AS-DP-Reset} show the execution time division of the {\tt Reset} state, where {\tt Cache} refers to {\tt DecodeCache} accesses. The {\tt DecodeCache} is an internal data structure of the decoder, implemented using an address map, which stores entries in page chunks. The {\tt DecodeCache} is accessed without making a memory request to \emph{Ruby} since it is part of the decoder. {\tt Bitfield} refers to bit field operations inside the function.

{\tt ModRM} will access attributes of the extended machine instruction (emi)---determine the next state based on its data and consume its ModRM byte---thus, it is likely that increasing memory will make it more likely for emi data to stay cached. The other FSM states are generally unaffected by more memory. Since the {\tt process opcode} state may transition to either {\tt ModRM} or {\tt Immediate}, where the former condition is always checked first, it is highly likely that the instructions dominating the applications (\textit{e.g.} {\tt Load}s) are prone to include {\tt ModRM} bytes. There are two possible reasons then why {\tt ModRM}'s successor states, \textit{e.g.} {\tt Immediate} may be less memory sensitive because few instructions require them.

Lastly, increasing the number of cores has comparatively less impact on the execution time division due to the FSM's bottlenecks being memory-related and owing to the fact that FSM are inherently sequential.
\profilingFigs{AtomicSimpleCPU}{AS-DP-Reset}{Execution division of the {\tt Reset} state}

\section{Execute}
\label{section:execute}
Each x86 instruction executed during the simulation will have its own root node under the AS CPU's \emph{tick}---there is not single {\tt execute} parent node since {\tt execute} is a virtual function that is overwritten depending on which x86 instruction is executing. This is unlike the other stages who do have a common root. The execution time division of each instruction type is detailed in Figures \ref{fig:AS-inst_gapbs}, \ref{fig:AS-inst_parsec3} and \ref{fig:AS-inst_spec2017}. There is great diversity between and within benchmarks, each necessitating its own set instruction set. Across the board, Load instructions (e.g. {\tt Ld}) of various types contitute a majority for most applications. Applications comparatively more dominated by Stores (e.g. {\tt St}) may dedicate more than a third of its execution time to those instructions.

We categorise the x86 instructions as specified in Table \ref{tab:ins_categories} due to the sheer instruction diversity. \textbf{Read memory} and \textbf{Write memory} instructions make up the bulk of the execution time. We can see in \profilingRefs{AS-exec-Ldbig} and \profilingRefs{AS-exec-St} how they heavily rely on \emph{Ruby}. We examine \emph{Ruby} in Section \ref{sec:AS_res_ruby}. Each instruction has a set of shared child function calls with its category peers, also shown in Table \ref{tab:ins_categories}.
\profilingFigs{AtomicSimpleCPU}{AS-inst}{Instruction sets of each application}
\begin{table}[htbp]
  \centering
  \small
  \begin{tabular}{| c || c |}
    \hline
    \textbf{Category} & \textbf{Operations} \\
    \hline
    \underline{\emph{Instruction pointer}} &  \\
    {\tt Wrip}, {\tt WripFlags}, {\tt WripImm}, {\tt Rdip} & {\tt getRegOperand}, {\tt readMiscRegOperand}\\
    \hline
    \underline{\emph{Write}}  & {\tt writeMemAtomic}, {\tt getRegOperand}, \\
     {\tt St}, {\tt Stul}, {\tt Stfp}     & {\tt readMiscRegOperand} \\
    \hline
    \underline{\emph{Read}} & {\tt readMemAtomic},{\tt setRegOperand}, \\
    {\tt Ld}, {\tt LdBig}, {\tt LdstBig}, {\tt Ldfp} & {\tt getRegOperand},{\tt readMiscRegOperand}\\
    \hline
    \underline{\emph{Logical}} & {\tt getRegOperand}, {\tt setRegOperand},\\
    {\tt Mxor}, {\tt XorFlagsBig}, {\tt AndFlagsBig}, {\tt XorFlags}, {\tt AndFlags}& {\tt setReg}\\
    \hline
    \underline{\emph{Arithmetic}} & \\
    {\tt Madf}, {\tt Mmulf}, {\tt Cvtf2f}, {\tt Cda}, {\tt SrlFlagsBig} & \\
    {\tt SrlFlagsImmBig}, {\tt SraFlagsImmBig}, {\tt SextFlagsImmBig}, & {\tt getRegOperand}, {\tt setRegOperand}\\
    {\tt SextFlagsBig}, {\tt AddFlagsBig}, {\tt AddFlagsImmBig}, & \\
    {\tt SubFlagsBig}, {\tt SubFlags}, {\tt SubImmBig}, {\tt Lfpimm}, {\tt Limm}, {\tt Lea} & \\
    \hline
    \underline{\emph{Move}} &  \\
    {\tt Mov}, {\tt MovFlags}, {\tt MovFlagsImm}, & {\tt getRegOperand}, {\tt setRegOperand}  \\
    {\tt BrFlags}, {\tt Mov2fp} & \\
    \hline
  \end{tabular}
  \caption{Categorisation of the x86 instructions used by the benchmarks.}
  \label{tab:ins_categories}  
\end{table}
\subsubsection{Read and write memory}
\label{sec:res-readmem}
Upon executing a Load or Store instruction, the CPU invokes \emph{read memory atomic} ({\tt readMemAtomic}) or \emph{write memory atomic} ({\tt writeMemAtomic}), respectively. This in turn issues a \emph{read}/\emph{write memory} operation; sending a packet to \emph{Ruby}'s memory response port. These memory access operations are by far the most dominant instruction as shown in \profilingRefs{AS-exec-Ldbig}, and \profilingRefs{AS-exec-St}, which use {\tt Ldbig} and {\tt St} as respective examples of Load and Store instructions.
\profilingFigs{AtomicSimpleCPU}{AS-exec-Ldbig}{Execution division for issuing a {\tt Ldbig}}
\profilingFigs{AtomicSimpleCPU}{AS-exec-St}{Execution division for a {\tt St} instruction}
\profilingRefs{AS-exec-ReadMem} show the execution time division for a \emph{read memory} operation of the AS CPU and \profilingRefs{AS-exec-WriteMem} show for \emph{write memory}. We clearly see that these dominant instructions are in turn spending the bulk of their time inside of \emph{Ruby}. We examine \emph{Ruby} in greater detail in Section \ref{sec:AS_res_ruby}.
\profilingFigs{AtomicSimpleCPU}{AS-exec-ReadMem}{Division to issue \emph{read memory} operation}
\profilingFigs{AtomicSimpleCPU}{AS-exec-WriteMem}{Division to issue \emph{write memory} operation}

\section{Post-execute}
\label{section:postexecute}
Once an instruction is finished executing, the Atomic CPU will perform the post-execute stage, updating internal statistics and notifying monitoring probes such as a performance monitoring units (PMUs). The internal statistics are logged to the CPU using the {\tt countCommitInst} function, and the simulator's probes are notified of the commit by the {\tt probeInstCommit} function. The division of the execution time inside post-execute is detailed in Figures \ref{fig:AS-PE_gapbs}, \ref{fig:AS-PE_parsec3} and \ref{fig:AS-PE_spec2017}. We group {\tt countCommitInst} and {\tt probeInstCommit} together under {\tt Commit}; {\tt Stat} are gem5's internal statistics; {\tt ThreadContext} are instructions related to obtaining the current thread's context; {\tt gem5-infra} refers to functions of gem5's infrastructure; {\tt C++} refers to operations done within C++. We observe that the majority of the post-execute stage is spent either updating these two statistics related to the aforementioned functions or for internal gem5 statistics ({\tt Stat}). Since these functions are closer to the leaf nodes of the function callgraph, pure {\tt C++} operations constitute a noticeable portion of their layers' execution.
\profilingFigs{AtomicSimpleCPU}{AS-PE}{Division of post-execute}
The partitioning of {\tt countCommitInst}'s execution time is detailed in Figures \ref{fig:AS-commitCPU_gapbs}, \ref{fig:AS-commitCPU_parsec3} and \ref{fig:AS-commitCPU_spec2017}. {\tt ThreadContext} are instructions related to obtaining the current thread's context. For each GAPBS application, increasing the memory from $3$GB to $8$GB consistently decreases the relative execution time of {\tt countCommitInst}. This is also the case the applications of PARSEC-3.0, where increasing from $3$GB to $16$GB reduces relative execution time for all applications except for {\tt raytrace}.
\profilingFigs{AtomicSimpleCPU}{AS-commitCPU}{Division of {\tt countCommitInst}}
The execution time division of {\tt probeInstCommit} is detailed in Figures \ref{fig:AS-pIC_gapbs}, \ref{fig:AS-pIC_parsec3} and \ref{fig:AS-pIC_spec2017} -- {\tt OpType} are instructions related determining the operation type of the finished instruction. Increasing the simulated memory decreases the relative time spent on {\tt C++} on GAPBS applications, whereas on PARSEC, increasing memory correlates to reducing the {\tt PMU} and {\tt OpType} portions and varying effect on {\tt gem5-Infra}. Increasing the number of simulated cores has a negligible impact across all applicatons.
\profilingFigs{AtomicSimpleCPU}{AS-pIC}{Division of {\tt countCommitInst}}
\paragraph{\textbf{\emph{Memory scaling's impact on post-execution}}:}
Increasing the number of cores has a minor impact on these functions---relative time spent inside post-execute is either slightly increased or decreased, depending on the application. Increasing the number of cores will not change the sequential nature of the CPU, which may be the reason why it has no major impact, however, adjusting the memory \emph{does} have an impact on both functions.
\begin{itemize}
\item For {\tt countCommitInst}, increasing the memory will decrease its relative execution time, across all applications. In PARSEC and GAPBS applications, we observe a correlation between increased memory and decreased relative time of {\tt C++} and {\tt gem5 infrastructure} ({\tt gem5-infra}) instructions. Otherwise, the pattern is that more memory correlates to proportionally less time spent inside post-execute.

\item For {\tt probeInstCommit}, increasing memory also correlates to less relative execution time. The trend across GAPBS applications is that the {\tt C++} portion is reduced, whereas for PARSEC applications, we instead observe that {\tt PMU}'s and {\tt OpType}'s portions are reduced, with varying effect on {\tt gem5-infra}. Like with {\tt countCommitInst}, increasing memory correlates to reduced {\tt gem5-infra} and {\tt C++} likely due to the decreased demand to simulate memory events, however, this function moreso experiences {\tt PMU} and {\tt OpType} instructions benefitting the most from more memory.
\end{itemize}

\section{\emph{Ruby}}
\label{sec:AS_res_ruby}
Figure \ref{fig:AS_anatomy} illustrates how a request at the \emph{Ruby}'s memory response is handled. To fetch an instruction or execute a {\tt Load}/{\tt Store}, the CPU sends an \emph{atomic request}, which is routed to the memory response port of \emph{Ruby}. Since all instructions are atomic in this CPU, each fetch request is handled by \emph{Ruby}'s receive Atomic ({\tt recvAtomic}) procedure. At \emph{Ruby}'s memory response port, it will validate the memory address: look inside the cache and see if it is mapped to physical memory. Once validated, assuming that it is not a PIO request, it is routed to \emph{Ruby}'s abstract controller which will handle the rest of the memory request. There, \emph{Ruby} will proceed after completing its access to the Abstract Memory class.

\profilingRefs{AS-Ruby-RR} show the execution time's partitions at the \emph{Ruby} memory response port. {\tt recvAtomic} denotes a \emph{receive atomic} request sent to the memory; {\tt mapAddrToMachine} denotes \emph{map address to machine}, which takes the instruction address and identifies its simulated device; \texttt{isPhysMemAddr} denotes \emph{is physical address}, which validates the address. We observe that for configurations running 3GB of simulated memory, {\tt isPhysMemAddr} takes up slightly more than half of the execution time, followed by {\tt recvAtomic}. Increasing the simulated memory from 3GB to 8GB in GAPBS or from 3GB to 16GB in PARSEC-3.0, correlates to increased time spent inside {\tt recvAtomic}, superseding {\tt isPhysMemAddr} as the most time-consuming function for these configurations.
\profilingFigs{AtomicSimpleCPU}{AS-Ruby-RR}{Division of \emph{Ruby} memory response port}
Figures \ref{fig:AS-RMR_gapbs}, \ref{fig:AS-RMR_parsec3} and \ref{fig:AS-RMR_spec2017} illustrate the runtime of the access to Abstract Memory. {\tt Addr} refers to address related operations, chiefly address validation; {\tt OpType} are the operations related to verifying the requested operation's type. {\tt Response} is related to creating a response used when the requester requires it. {\tt Write} and {\tt Read} are for write and read instructions, respectively. {\tt gem5-infra} refers to functions of gem5's infrastructure.

The access to the Abstract Memory is chiefly dominated by {\tt Addr}, with {\tt gem5-infra} taking a roughly equal amount of time. {\tt Read} is comparatively more prominent than {\tt Write} for most applications, which is intuitive since the results in Section \ref{section:execute} corroborates this. For example, {\tt cc\_raw} of GAPBS has comparatively more prominent {\tt Write} operations, which is consistent across the Figures for the memory access (Figure \ref{fig:AS-RMR_gapbs}) and the dynamic instruction (Figure \ref{fig:AS-inst_gapbs}).
\profilingFigs{AtomicSimpleCPU}{AS-RMR}{Division of accessing Abstract Memory}
\subsubsection{\textbf{MMU}}
The MMU will defer all requests to the TLB, which will invoke its {\tt translate} function and in turn the {\tt lookup} function. If it could not find an entry in {\tt lookup}, the TLB will invoke its Page Walker, which will retrieve the address via a \emph{Ruby} request. Therefore, we can claim that invoking the {\tt Walker} is synonymous with fetching data from \emph{Ruby}. \profilingRefs{AS-MMU} show the division of execution time at the TLB's {\tt translate} function.
\profilingFigs{AtomicSimpleCPU}{AS-MMU}{Division of the TLB's {\tt translate} function}
We could previously note that in Figure \ref{fig:AS-exec-ReadMem_gapbs} in GAPBS, {\tt bfs\_raw} relied heavily on the MMU to execute \emph{read memory} instructions, and Figure \ref{fig:AS-MMU_gapbs} reveals that this in turn is largely due to it needing to wait for the Walker, which employs \emph{Ruby}---\profilingRefs{AS-MMU-Walker} show that the Walker \emph{does} spend most of its time waiting on \emph{Ruby}, \textit{i.e.} the true performance bottleneck for applications such as {\tt bfs\_raw} is not the MMU itself, but rather \emph{Ruby}, as is the case with the all other applications.
\profilingFigs{AtomicSimpleCPU}{AS-MMU-Walker}{Division of the TLB's Walker}
\subsubsection{\textbf{\emph{Ruby} as a bottleneck}}
As discussed in Sections \ref{section:fetchins} and \ref{section:execute}, fetching an instruction and executing a {\tt Load} or {\tt Store}, spend the vast majority of their execution time inside \emph{Ruby}.

We observe in Figures \ref{fig:AS-overall_gapbs}, \ref{fig:AS-overall_parsec3} and \ref{fig:AS-overall_spec2017} that all applications except for {\tt lbm}, are \emph{instruction fetch}-bound, regardless of their configuration. An AS CPU executes instructions sequentially without pipelining, and therefore should not inherently encounter significant memory latency. Nevertheless, \emph{Ruby} is demonstrably a bottleneck for these configurations---it introduces overhead, even in single-core configurations. Figure \ref{gprof_60min_callgraph_4p}, generated with gprof data, supports this notion too as all functions taking up \textbf{less than 4\% of the total runtime} have been removed from the graph, and solely \emph{Ruby} remains. Furthermore, we observe in \profilingRefs{AS-inst} that regardless of specific instruction set, all applications spend the most time on {\tt Load}/{\tt Store}, which also access \emph{Ruby}; outliers such as {\tt lbm} are still highly likely to be \emph{Ruby}-bound.

{\tt tc\_raw}, \textit{i.e.} \emph{triangle counting} and {\tt bc\_raw}, \textit{i.e.} \emph{betweeness centrality} both exhibit more complex control flows involving nested graph traversals, which correlate with higher demand to fetch instructions comparatively greater than their GAPBS counterparts.

A trend observed on all applications is that increasing memory will increase the proportional time spent in {\tt fetch}, albeit with a more noticeable difference in the aforementioned GAPBS applications. Inside of the \emph{Ruby} operations, (see Figures \ref{fig:AS-Ruby-RR_gapbs}, \ref{fig:AS-Ruby-RR_parsec3}) we observe that increasing memory will \emph{relatively decrease} time spent inside {\tt isPhysMemAddr}, but \emph{relatively increase} time spent on {\tt recvAtomic}. Since the increased time in {\tt recvAtomic} is relatively greater than the decreased time spent on {\tt isPhysMemAddr}, the total relative execution time is \emph{increased} with more memory at the \emph{Ruby} memory response port.

\emph{Ruby}'s overhead may be caused by increased complexity of address translation with greater memory, as address mappings and routing will traverse a greater space. We can observe in Figures \ref{fig:AS-RMR_gapbs} and \ref{fig:AS-RMR_parsec3} that access to memory constitutes a comparatively lesser portion with greater memory. Nevertheless, \emph{Ruby}'s overhead negates performance benefits from increasing memory, as observed on higher levels than the memory access.

The impact of increasing the number of cores is not as impactful; \textit{e.g.} the impact on \emph{Ruby}'s memory response port is negligible.

\chapter{Results for Timing Simple CPU}
\label{sec:TS_CPU_breakdown}
\begin{figure}[p]
  \centering
      \rotatebox{90}{
\includegraphics[width=1.4\textwidth]{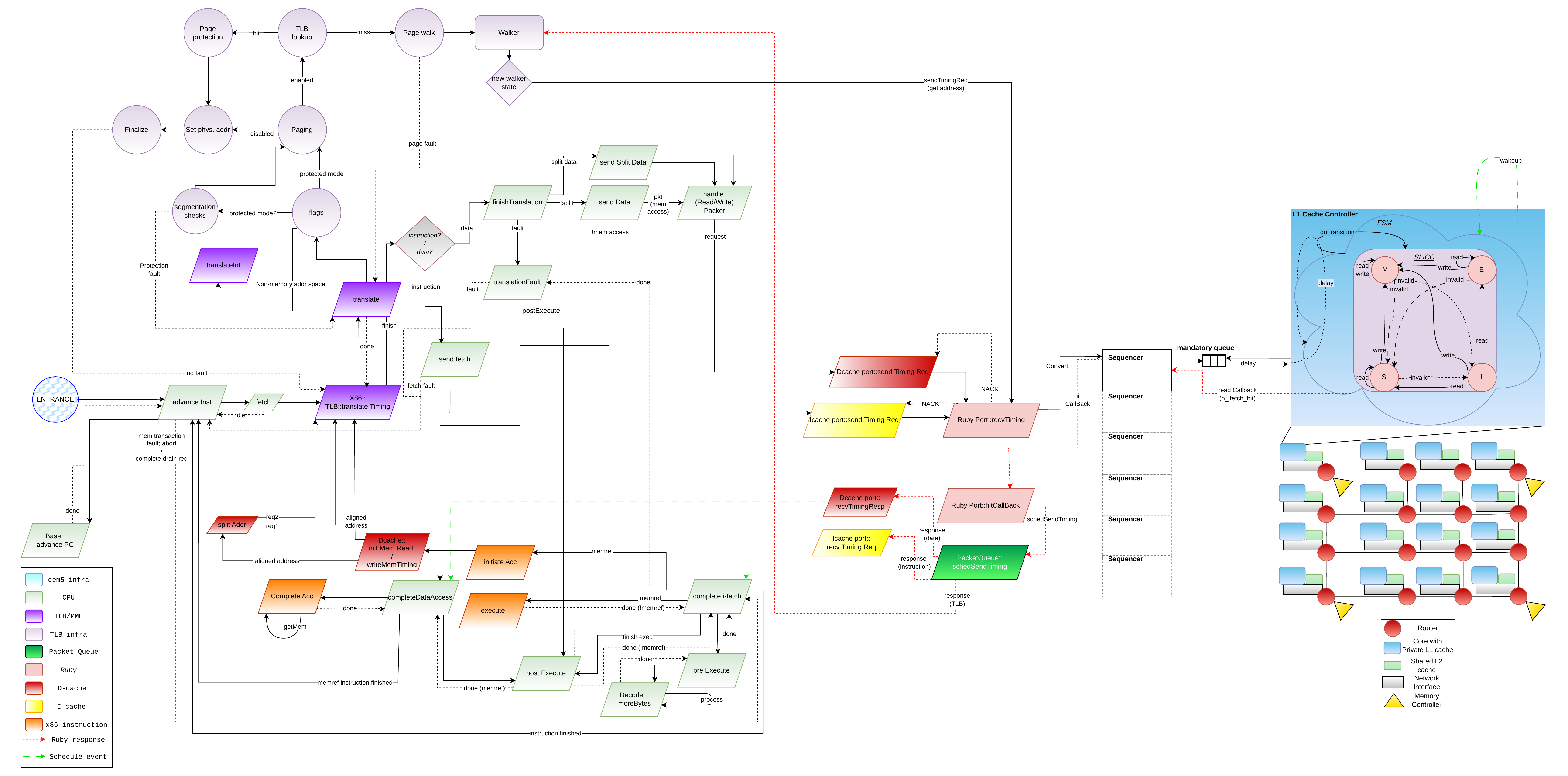}
}
\caption{\label{fig:TS_anatomy} Anatomy figure showing the function call interactions between gem5's components running the TS CPU.}
\end{figure}

The top-level of the TimingSimpleCPU's (TS CPU) execution consists of chiefly three {\tt process} function calls: 1) a wrapper for \emph{Ruby}, 2) \emph{tick} of the data cache ({\tt D-tick}), and 3) \emph{tick} of the instruction cache ({\tt I-tick}). The \emph{Ruby} wrapper is used for memory system events, such as memory accesses. {\tt D-tick} is triggered upon data cache accesses---related to \textit{e.g.} load and store instructions, and {\tt I-tick} is triggered upon an instruction fetch. We show a detailed anatomy of the TS CPU's execution flow in Figure \ref{fig:TS_anatomy}. It includes a legend explaining the colour coding. Dashed arrows are used to inidicate responses and functions returning to their caller. For example, the red dashed arrows denote the path of a \emph{Ruby} response. We examine this control flow in greater detail in subsequent sections, in conjunction with presenting the profiling results.
\profilingRefs{TS-overall} show the division of execution time for the top-level function. We can clearly observe how \emph{Ruby} dominates across all GAPBS applications, \textit{e.g.} {\tt cc\_raw} having overall more time spent inside {\tt I-tick} than its peers. This is also the case for certain applications of PARSEC-3.0 and SPEC2017, where {\tt I-tick} and {\tt D-tick} are comparatively more prominent. For instance, {\tt blackscholes} of PARSEC spend the majority of its execution time inside {\tt I-tick} unlike {\tt dedup}, which is \emph{Ruby}-bound.
\profilingFigs{TimingSimpleCPU}{TS-overall}{Division of overall run-time}
We also examine this layer from a \emph{flattened} perspective, \textit{i.e.} characterising execution time partioning in terms of bottommost legible layer. \profilingRefs{TS-flatten} show the operations previously hidden by underlying functions that we spend time inside. We observe that {\tt Ruby} and its infrastructure ({\tt Ruby-Infra}) are collectively the most time-consuming portion for all applications. We observe no significant impact of increasing memory of the number of cores on this layer of execution. 
\profilingFigs{TimingSimpleCPU}{TS-flatten}{Division of overall \emph{flattened} run-time}
\profilingRefs{TS-itick} show that {\tt I-tick}'s execution time is almost solely dedicated to {\tt complete i-fetch}, which is the callstack's top-function for fetching instructions. This is because {\tt complete i-fetch} is the \emph{de facto} entrance point of {\tt I-tick}. However, its first call will not process an instruction as it will instead go straight to {\tt advance Inst}, thus starting the greater procedure. We examine this in Section \ref{TS_res_fetch}. 
\profilingFigs{TimingSimpleCPU}{TS-itick}{{\tt I-tick}'s division of execution time}

\section{Instruction fetch}
\label{TS_res_fetch}
We illustrate how an instruction is fetched from memory in the TS CPU in Figure \ref{fig:TS_anatomy}: light green boxes denote CPU-side operations; violet denotes the translation lookaside buffer (TLB); light yellow denotes instruction cache ({\tt i-cache}) and light red denotes \emph{Ruby}.

The CPU will call {\tt advance Inst} ({\tt advance Instruction}) to advance its PC \emph{if} the CPU is running; otherwise it is stalled and must schedule an event and try again---as is the case for several stages of the execution flow. If allowed to advance, {\tt fetch} will set up a memory request, subsequently sent to the TLB. If the address is present in the TLB, we immediately proceed; otherwise, we must perform a page walk and wait. We show the TLB's execution time division in Section \ref{sec:TS_TLB}. Once finished, the TLB will respond back to the CPU, triggering its {\tt send fetch} function, sending a packet to the instruction cache. The instruction cache will construct a {\tt Timing Request}, which will be sent to \emph{Ruby}'s memory response port. In particular, the port's {\tt recv} ({\tt receive}) {\tt Timing Request} function. \emph{Ruby} will then fetch the instruction from memory, and callback once finished.

Once the instruction is ready to execute, an event for {\tt complete i-fetch} is scheduled and we therefore re-enter it to proceed with instruction execution. This event will appear as a child of the CPU in the callstack. However, it is not recursion, rather the event-driven nature of the TS CPU making it appear so in the callstack. This allows us to differentiate this call to {\tt complete i-fetch} from the one at the {\tt I-tick} process level\footnote{As examined in the overview section and demonstrated in \profilingRefs{TS-itick}.}. Therefore, we can examine the execution of an instruction as this separate instance of {\tt complete i-fetch} in the callchain---we show those results in Section \ref{TS_res_exec}. {\tt complete i-fetch} calls upon {\tt advance Inst} to proceed to the next instruction. We clearly see in \profilingRefs{TS-adinst} that {\tt advance Inst} is solely dedicating its runtime towards {\tt fetch}.
\profilingFigs{TimingSimpleCPU}{TS-adinst}{Division of run-time for {\tt advance Inst}}

The execution time for {\tt fetch} is shown in \profilingRefs{TS-fetch-overall}. {\tt send\_fetch\_request} encompasses sending and receiving a response by \emph{Ruby} for the instruction fetch request.\\ {\tt advanceInst-[ROM/S-Inst]} denotes executing \emph{Static} instructions or accessing \emph{read-only memory} (ROM); thus not requiring to send a request to \emph{Ruby}. {\tt TLB::translateTiming} is the parent function of the TLB, which performs the lookup. It is called by {\tt BaseMMU::translateTiming}. Once the TLB has found the address, it will call upon {\tt FetchTranslation::finish}.
\profilingFigs{TimingSimpleCPU}{TS-fetch-overall}{Division of overall run-time for {\tt fetch}}
\profilingRefs{TS-fetch} show a more fine-grained categorisation of {\tt fetch}.
\profilingFigs{TimingSimpleCPU}{TS-fetch}{Fine-grained time division for {\tt fetch}}
The impact of increasing the number of cores or memory is inconclusive as shown in \profilingRefs{TS-fetch-overall}. For instance, {\tt bfs\_raw}, {\tt cc\_sv\_raw} and {\tt pr\_spmv\_raw} and {\tt tc\_raw} of GAPBS will be relatively unaffected with the notable exception being for $4$ cores and $3$GB, which correlates with comparatively longer time spend inside the instrucion fetch stage. For PARSEC, most applications are generally unaffected by varying the configurations with a notable exception being {\tt facesim}, which spends comparatively longer time on fetching an instruction while running $4$ cores with $16$GB and $16$ cores with $16$GB.

\section{Execute}
\label{TS_res_exec}
Upon receiving the response for the instruction fetch from \emph{Ruby}, the event scheduler will schedule {\tt complete i-fetch} to begin the decoding, execution, and post-execution stages. It will begin with calling {\tt pre Execute} to perform instruction decoding. We then proceed with the execution itself. The instruction execution control flow is illustrated in Figure \ref{fig:TS_anatomy}---{\tt complete i-fetch} invokes the instruction's {\tt execute} function if it will not reference memory; otherwise it will invoke its {\tt initiate Acc} ({\tt initiate Access}) procedure. Orange boxes denote x86 instruction interface and red denotes the data cache. 

If the instruction does not contain a {\tt memRef}, we simply execute it, proceed to {\tt post Execute}, \textit{i.e.} committing, and finish. If we do have {\tt memRef}, {\tt initiate Acc} will need to fetch the physical address from the TLB via the data cache. It invokes a function of the data cache; {\tt init Mem Read} or {\tt write Mem Timing}, depending on if the instruction is a Load or Store, respectively. This will request an address translation from the MMU and its TLB, performing the \emph{translate Timing} procedure. If the TLB misses with its lookup, it must perform a page walk, which will fetch the desired physical address via \emph{Ruby}. We show the execution time divison of this procedure in Section \ref{sec:TS_TLB}.

The TLB will eventually complete its inquiry and invoke {\tt finishTranslation}, which will send the \emph{Ruby} data request for instruction execution. Once \emph{Ruby} is finished, it will schedule a response that will arrive at the D-cache's {\tt recvTimingResp} port. This will invoke {\tt completeDataAccess}, which will run the instruction's {\tt Complete Acc}---thus completing the memory access.

Once the instruction's execution is done, we enter {\tt post Execute}, which commits the instruction. Afterwards, we enter {\tt advance Inst} to proceed to the next instruction.
\profilingRefs{TS-Cmpl-ifetch} show the execution time division of {\tt complete i-fetch} at the TS CPU level; the top-level function for instruction execution. We can clearly see that initiating and completing a memory reference instruction, denoted as {\tt DCache-Access[MMU]}, and executing it ({\tt D-inst}) is insignificant compared to {\tt advInst}, which is equivalent to instruction fetching from this layer.

{\tt advInst} is consistently the most time-consuming aspect of instruction execution in the TS CPU. This procedure will in turn need to rely upon \emph{Ruby} for address translation and memory access as explained in Section \ref{TS_res_fetch}. The address translation and the data access are driven by the TLB for both the {\tt I-cache} and the {\tt D-cache}. 
\profilingFigs{TimingSimpleCPU}{TS-Cmpl-ifetch}{Run-time division for {\color{violet}{\tt complete i-fetch}}}
\section{TLB}
\label{sec:TS_TLB}
The TS CPU will invoke the TLB both during the \emph{instruction fetch} and \emph{execute} stages. This is also denoted as \emph{MMU} in some figures, but we know that the {\tt translateTiming} function of the MMU will directly defer the request to the TLB. The TLB routes the request to its {\tt translate} function, which in turn will call upon {\tt lookup} to see if the provided virtual address has a valid TLB entry. If {\tt lookup} hits, it will set the physical address and proceed; otherwise if it misses, the TLB will invoke its {\tt Walker} and perform a Page Walk to obtain the physical address. We show the division of execution time of {\tt translate} in \profilingRefs{TS-MMU}---{\tt TLB-infra} refers to other operations of the TLB. We clearly observe that it spends most of its execution time inside the {\tt Lookup} function and that certain applications tend to trigger the {\tt Walker} more than others, such as {\tt bc\_raw}, {\tt bfs\_raw} and {\tt tc\_raw} of GAPBS and {\tt canneal} of PARSEC.

\profilingFigs{TimingSimpleCPU}{TS-MMU}{Run-time division of TLB's {\tt translate}}
The Walker will obtain the desired physical address by sending a request to \emph{Ruby}. We show the division of execution time for the Walker in \profilingRefs{TS-TLB-Walker}, where {\tt walker-Infra} denotes functions related to its infrastructure. Across all applications that do prominently invoke the Walker, we observe that it will spend the vast majority of its execution time waiting for \emph{Ruby}.
\profilingFigs{TimingSimpleCPU}{TS-TLB-Walker}{Run-time division of TLB's Walker}
\section{\emph{Ruby}}
\label{TS_res_ruby}
As demonstrated in \profilingRefs{TS-flatten} and explained in Sections \ref{TS_res_fetch} and \ref{TS_res_exec}, \emph{Ruby} constitutes a significant portion of the total execution time, and is the single most time-consuming component of the TS CPU once we also include its entire infrastructure.

\profilingRefs{TS-ruby} show the execution time partitions of \emph{Ruby} at its top-level {\tt processCurrentEvent} function. To ensure \emph{timed memory accesses}, events are enqueued and eventually processed at this level of the callstack. We observe that Garnet dominates the execution time for certain applications, namely all in GAPBS, {\tt canneal}, {\tt dedup} and {\tt facesim} of PARSEC, {\tt bwaves}, {\tt lbm} and {\tt xalancbmk} of SPEC. Whereas the rest are dominated by the L1 cache. Referring to Figure \ref{fig:TS_anatomy}, this section covers the interaction between the Sequencer and the \emph{Ruby} port and the internal operations of the L1 cache controller (L1 CC), including its FSM. We present the results related to Garnet in Section \ref{TS_res_garnet}. This layer reveals how there are three major contenders for being most time-consuming: {\tt garnet-Router}, {\tt garnet-NI} or the {\tt L1Cache}. Generally, {\tt garnet-Router} tends to dominate the execution time for GAPBS and certain PARSEC applications ({\tt dedup}, {\tt facesim}), and {\tt L1Cache} dominates in several PARSEC applications: {\tt blackscholes}, {\tt bodytrack}, {\tt ferret}, {\tt fluidanimate}, {\tt freqmine}, {\tt raytrace}, {\tt swaptions} and {\tt vips}. SPEC is more diverse, though {\tt L1Cache} is generally the most time-consuming, with exception applications being {\tt bwaves} and {\tt lbm}, who are instead spending the most time inside {\tt garnet-Router} and {\tt garnet-other}.

Figure \ref{fig:AS-inst_parsec3} reveals that instruction set diversity alone is not determining whether the application will spend more time inside the {\tt L1Cache} or {\tt garnet-Router}. It is rather application-dependent, where factors such as branching or memory-dependencies, will determine whether an application is prone to hitting (most time on {\tt L1Cache}) or missing in the L1 cache (most time on {\tt garnet}).
\profilingFigs{TimingSimpleCPU}{TS-ruby}{Division of execution time of \emph{Ruby}}
The Sequencer acts as the interface between the \emph{Ruby} port and the greater memory system. A request is sent through the Sequencer; the response is likewise sent from the Sequencer in a {\tt hitback} call. \profilingRefs{TS-L1-RHC-overall} illustrate the overall division of execution time for such {\tt hitback} calls. We clearly observe that the majority of execution time is spent on waiting for a response, as denoted by the {\tt send\_I/D\_back} category. The remaining categories pertain to event scheduling---{\tt EventQueue} and {\tt EventManager} through which requests and responses are scheduled.
\profilingFigs{TimingSimpleCPU}{TS-L1-RHC-overall}{Hitback call's execution time division}
\profilingRefs{TS-L1-RHC} present a more fine-grained overview of the layer. They show that the slowest steps of the request-response procedure are {\tt schedTimingResp}, {\tt schedSendTiming} and {\tt schedSendEvent}---together they schedule and manage sending a timed package. More specifically, {\tt schedTimingResp} schedules a timing response packet, {\tt schedSendTiming} adds a packet to the transmit list, scheduling it, and {\tt schedSendEvent} schedules a \emph{send} event.
\profilingFigs{TimingSimpleCPU}{TS-L1-RHC}{Hitback call's fine-grained execution time}
\profilingRefs{TS-L1Cache} show the execution time of the L1 CC's FSM changing states. We clearly observer across all applications, that the vast majority of time is spent on executing the state task ({\tt TransWorker (L1State)}).
\profilingFigs{TimingSimpleCPU}{TS-L1Cache}{L1 cache FSM operations' execution time}
\profilingRefs{TS-L1CC} show the overall actions of the L1 CC's actions and their execution time division. {\tt store-hit} are store operations that hit in the cache; {\tt load-hit} is the same but for load operations. {\tt pop-man} is dequeuing a request from the \emph{mandatory queue} in the L1 CC. {\tt ifetch-hit} denotes an instruction fetch hit in the cache. 
\profilingFigs{TimingSimpleCPU}{TS-L1CC}{Overall division of L1 CC actions}
\profilingRefs{TS-L1State} show the execution time division at the same layer, however more fine-grained. We configured \emph{Ruby} to run a two-level MESI protocol, with these figures showing the actual actions of the gem5 implementation of the protocol. We clearly observe that the following subset of actions are particularly notable in terms of consuming execution time:
\begin{itemize}
\item {\tt h\_load\_hit}: a load request hits in the L1 D-cache; notify the Sequencer.
\item {\tt h\_ifetch\_hit}: an instruction request hits in the L1 I-cache; notify the Sequencer.
\item {\tt hx\_load\_hit}: load request hit; notify the Sequencer and update most recently used (MRU) of the L1 I-cache and D-cache.
\item {\tt hh\_store\_hit}: store request hit; notify the Sequencer and update the D-cache MRU.
\item {\tt hhx\_store\_hit}: store request hit; notify the Sequencer and update the MRU of the D-cache and I-cache.
\item {\tt k\_popMandatoryQueue}: dequeue request from mandatory queue.
\item {\tt g\_issuePUTX}: evict a modified or exclusive cache line and writeback it to the L2 cache.
\item {\tt forward\_eviction\_to\_cpu}: evict clean cache line from the L1 cache; notify the CPU via the Sequencer.
\end{itemize}
Of these, {\tt h\_ifetch\_hit} tends to take up a noticeable portion of the execution time for all applications. Taking the Garnet figures in Section \ref{TS_res_garnet} into account, we observe that applications that tend to miss in the L1 cache, \textit{i.e.} spend proportionally more time inside Garnet, still spend the most time inside {\tt h\_ifetch\_hit}, albeit it overall will spend less time inside the L1 CC altogether compared to appplications that tend to hit in the L1 cache.
\profilingFigs{TimingSimpleCPU}{TS-L1State}{L1 CC's actions' execution time division}

\section{Garnet}
\label{TS_res_garnet}
The wake-up function to the router is at the top-level of Garnet, which in turn calls upon its switch allocators and processes the router's pipeline each cycle. We show the execution time division for this level in \profilingRefs{TS-Garnet-Router}. A {\tt flit} is the data packet unit, and {\tt credits} denote a router's capacity. The router wake-up function will wake-up its {\tt flits} and {\tt credits} to process incoming data and buffer capacity, respectively. Afterwards, it will invoke the {\tt SwitchAllocator} to manage the router's data flow.

We summarise the categories of \profilingRefs{TS-Garnet-Router} below:
\begin{itemize}
\item The router wake-up function will begin by calling wake-up on all incoming {\tt flit}s; invoking {\tt InputUnit::wakeup}, which calls {\tt InputUnit::route\_compute} on the head {\tt flit}\footnote{The head {\tt flit} denotes the first {\tt flit} of a data packet.}. {\tt route\_compute} calls {\tt RoutingUnit::outportCompute}, which determines the suitable output port for the current {\tt flit}. It will consult the \emph{routing table} with\\ {\tt RoutingUnit::lookupRoutingTable} for this task. This involves using\\ {\tt NetDest::intersectionIsNotEmpty} to see if an output link overlaps with a {\tt flit}'s destination.
\item A {\tt flitBuffer} contains {\tt flit}s as they await being processed by a router. It is marked as ready in {\tt flitBuffer::isReady} if the buffer is not empty and has a top {\tt flit} that is currently ready. This function is invoked at numerous stages in the router.
\item The {\tt wakeup} call of the {\tt SwitchAllocator} will invoke its 2-stage process of managing data flow. (1) {\tt arbitrate\_inports} will in a round robin manner select one virtual channel per input port that can request an output port. (2) {\tt arbitrate\_outports} will also in a round robin manner, go through all output ports and select a ``winner'' input port, thereby scheduling its {\tt flit} to traverse through. {\tt arbitrate\_outports} then updates the credits the corresponding virtual channel with {\tt InputUnit::increment\_credit}, to signal that a slot has been freed. The {\tt Credit::Credit} constructor is then used to generate a new {\tt credit} object that is propagated back to the VC.

\item The {\tt SwitchAllocator}'s {\tt wakeup} is then concluded by calling {\tt check\_for\_wakeup} which schedules a wake-up for each {\tt flit} that needs one. This is checked using {\tt need\_stage} -- denoted by the {\tt InputUnit::need\_stage} category.
\end{itemize}

\profilingFigs{TimingSimpleCPU}{TS-Garnet-Router}{Run-time division of Garnet's router}

\profilingRefs{TS-Garnet-NI} show the exeuction time of Garnet's network interface's wake-up function. The wake-up function will check for ready messages in its protocol buffer, split them into {\tt flit}s, place them into an output buffer, and schedule their output links. Additionally, incoming {\tt flit}s in its input links are picked up and examined---reassembled into a message if the current {\tt flit} was a tail; otherwise, it notifies the sender with a {\tt credit}.  

We summarise each category in \profilingRefs{TS-Garnet-NI} below:
\begin{itemize}
\item {\tt NI::flitisizeMessage} is used to convert a message into {\tt flit}s. During this process, {\tt RouteInfo::RouteInfo} constructors are created to store the route into each {\tt flit}. Similarly, {\tt NetDest} contains the message's destination---{\tt NetDest::NetDest} denotes its constructor which invokes {\tt NetDest::resize} for setting up bitmasks for each machine type to later mark each valid destination.
\item {\tt NI::scheduleOutputlink} is called to examine NI buffers for ready {\tt flit}s. It invokes {\tt NI::scheduleOutputPort} for each output port to select its virtual channel and checks if it has ready data, while also ensuring it is correctly ordered and has sufficient credit. If that is the case then the virtual channel is selected and we decrement its {\tt credit} ({\tt Credit::Credit}) and schedule the {\tt flit}.
  
\item Lastly, {\tt NI::checkReschedule} is invoked to wake-up the NI in the next cycle if there are waiting messages in the protocol buffer or {\tt flit}s in the VC buffer. 
\end{itemize}
\profilingFigs{TimingSimpleCPU}{TS-Garnet-NI}{Network Interface's run-time division}
As shown, it is difficult to single out a bottleneck operation within Garnet as the execution time division appears to be more or less evenly distributed. We do, however, reveal which applications and configurations spend more time inside Garnet. Since Garnet is invoked every time a request misses in the L1 cache, if we cross-reference our finding with the figures in Section \ref{TS_res_ruby}, we can observe a correlation between an application's tendency to spend more time inside Garnet and therefore less time inside the L1 CC when compared to an application that tends to \emph{hit}.

\chapter{Results for the O3 CPU}
\label{sec:O3_CPU_breakdown}
Because of the O3 CPU’s complex execution flow, we present each stage in separate anatomical overview figures. Stages do not communicate with one another directly, but through the \texttt{TimeBuffer} objects in gem5. A TimeBuffer is a circular buffer that models inter-stage communication by enforcing latency: signals issued by one stage become visible to the next only after a set number of CPU cycles. For instance, the toIEW TimeBuffer delivers signals from the rename stage (the upstream stage) and commit stage (the downstream stage) to the IEW stage after 1 simulated clock cycle by default.

The O3 CPU will periodically invoke a series of {\tt tick}s in the following order: (1) {\tt Fetch::tick}, (2) {\tt Decode::tick}, (3) {\tt Rename::tick}, (4) {\tt IEW::tick} and (5) {\tt Commit::tick}. Afterwards, it will call {\tt TimeBuffer::advance} to advance the time in the simulator. We present the execution time partitions of the parent {\tt tick} function in \profilingRefs{O3-core}. Clearly, the IEW stage takes up most execution time for all applications, with the Fetch stage taking up the second most time.
\profilingFigs{O3CPU}{O3-core}{Division of execution time at O3 CPU's top level}

\profilingRefs{O3-overall} show the \emph{overall} execution time of the O3 CPU, each stage but also including top-level \emph{Ruby} function calls. \emph{Ruby} is examined in Section \ref{res_o3_ruby} and the Garnet router in Section \ref{res_o3_garnet}. At this level, applications either spend most execution time inside IEW or \emph{Ruby}.
\profilingFigs{O3CPU}{O3-overall}{Overall execution time division of the O3 CPU}
\section{Fetch}
\label{sec:o3Fetch}
\begin{figure}[H]
  \centering
      \rotatebox{90}{
\includegraphics[width=1.3\textwidth]{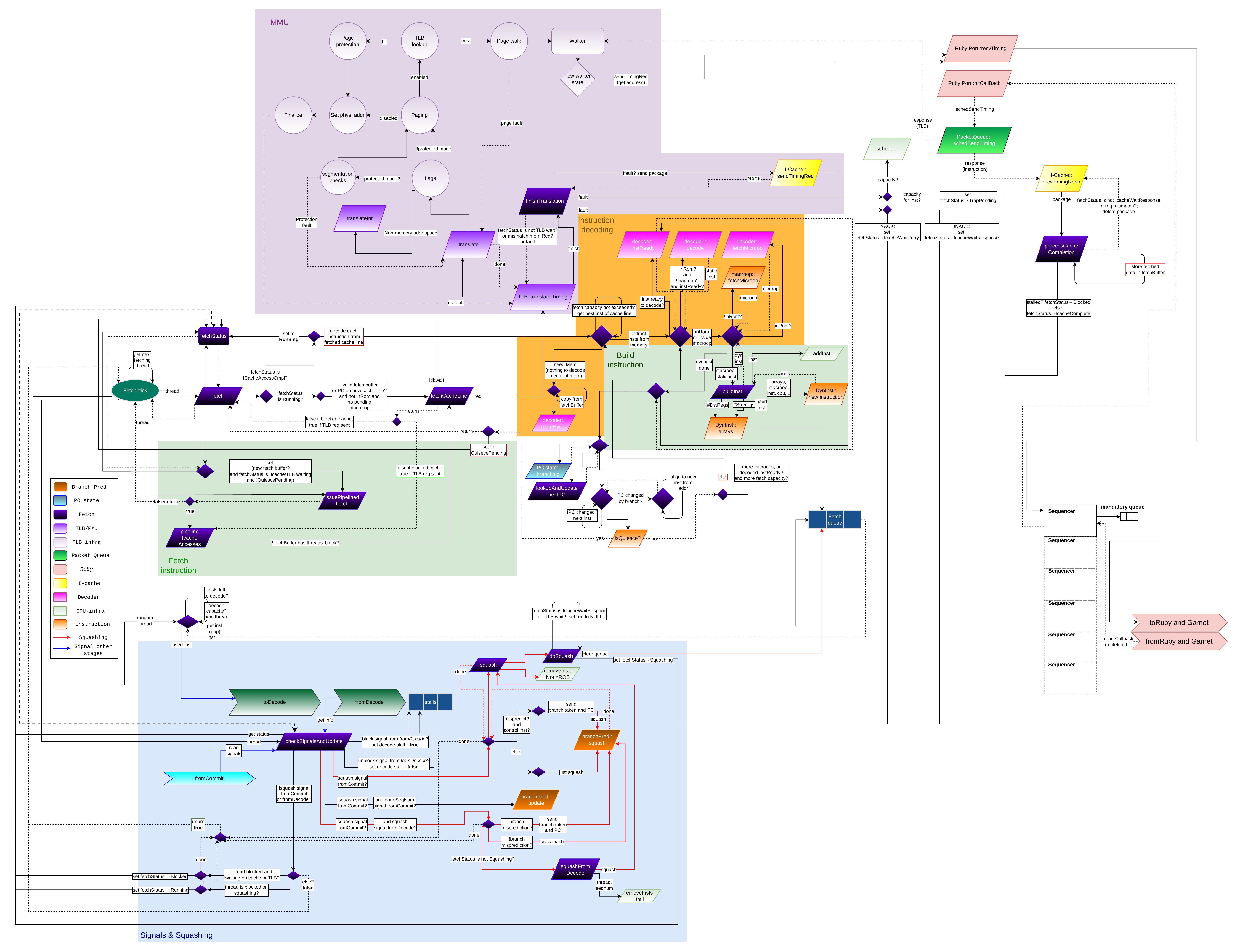}
}
      \caption{\label{fig:O3_Fetch} Anatomy figure showing the execution flow of the O3 CPU's Fetch stage.}
\end{figure}
We illustrate the execution flow of the Fetch stage in Figure \ref{fig:O3_Fetch}. At its top-level is the stage's {\tt tick} function---\profilingRefs{O3-fetch-tick} show its execution time division. For each thread, the Fetch stage first checks incoming signals from the Commit and Decode stages and update the Fetch stage's status based on their signals ({\tt checkSignal}). It will also squash instructions if squash signals are received from these stages. Afterwards, for each thread, it invokes the stage's core function {\tt fetch}. If needed, the core {\tt fetch} function will fetch the cache line containing the desired intruction; otherwise, if it has completed the instruction cache (I-cache) access, it will decode the instruction such that a {\tt DynInst} can be created and stored in the {\tt Fetch Queue}. Subsequently, the top-level {\tt tick} will attempt to pipeline I-cache requests ({\tt pipeline-Icache-req}) if allowed to; \textit{e.g.} if it is not currently waiting on the TLB. It will then send each previously stored instruction to Decode. 
\profilingFigs{O3CPU}{O3-fetch-tick}{Execution time division for Fetch's tick}
\profilingRefs{O3-fetch-tick-flatten} show the \emph{flattened} execution time of Fetch's {\tt tick} function. We observe that this reveals how all applications spend a significant time on construction an instruction at this stage, and comparatively little time on memory requests ({\tt pipelineIcacheAccesses}, {\tt fetchCacheLine}). 
\profilingFigs{O3CPU}{O3-fetch-tick-flatten}{Flattened execution time for Fetch's tick}
\profilingRefs{O3-fetch-inst} show the execution time division of the core {\tt fetch} function. The reason why {\tt Icache} is not a prominent portion of the execution time at this level is due to {\tt fetch}'s execution flow---if it needs to send a request to the I-cache, it will do so and then immediately return to its caller, the {\tt tick} function. If in a subsequent CPU {\tt tick}, the I-cache access has completed, {\tt fetch} can then proceed with the instruction decoding and construction; the actual time-consuming portions. We observe that the {\tt Decoder} makes up roughly less than half of the relative time compared to {\tt build-inst}, \textit{i.e.} the instruction construction.
\profilingFigs{O3CPU}{O3-fetch-inst}{Execution time division for the core {\tt fetch} function}
\profilingRefs{O3-fetch-buildInsts} show the execution time division for constructing/building an instruction. We observe that by far, most time is spent inside the dynamic instruction constructor itself ({\tt D-Inst-Constructor}).
\profilingFigs{O3CPU}{O3-fetch-buildInsts}{Execution time division for {\tt buildInst}}
The stage will fetch a cache line either inside the core {\tt fetch} function or pipelined in a subsequent function call to {\tt fetchCacheLine}. \profilingRefs{O3-fetch-pipeline-cl} show the flattened execution time division of {\tt pipelineIcacheAccesses}, clearly showing that the vast majority of its execution time is inside TLB-related functions.
\profilingFigs{O3CPU}{O3-fetch-pipeline-cl}{Flattened execution time of {\tt pipelineIcacheAccesses}}
\profilingRefs{O3-fetch-Signals} show the execution time division {\tt checkSignalsAndUpdate} where we see most time is spent on operations of the branch predictor ({\tt Branch-pred}) for most applications. A notable exception is {\tt dedup} of PARSEC, where {\tt squash} takes up most of its execution time for certain configurations. Furthermore, {\tt bc\_raw}, {\tt pr\_raw}, and {\tt pr\_spmv\_raw} of GAPBS, and {\tt bwaves}, {\tt cactuBSSN}, and {\tt lbm} of SPEC2017, along with  {\tt canneal} and {\tt facesim} of PARSEC, spend most of their time inside operations related to receiving signals from or sending signals to other stages ({\tt signals}). Overall, {\tt squash}'s and {\tt signals}' percentage depends on applications.
\profilingFigs{O3CPU}{O3-fetch-Signals}{Execution time division for Fetch's {\tt checkSignalsAndUpdate}}

\section{Decode}
\label{sec:o3Decode}
\begin{figure}[H]
  \centering
      \rotatebox{90}{
\includegraphics[width=1.3\textwidth]{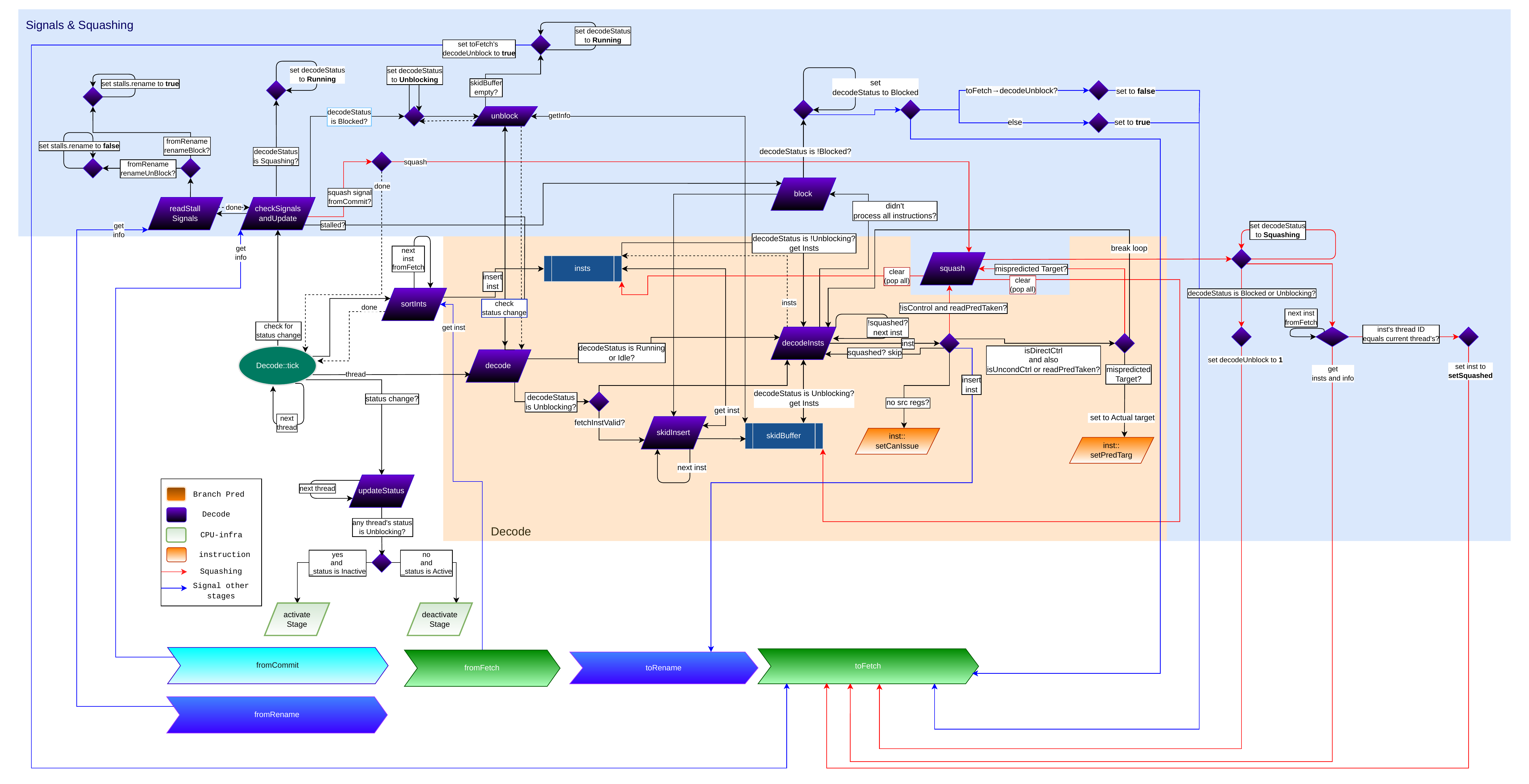}
}
\caption{\label{fig:O3_Decode} Anatomy figure showing the execution flow of the O3 CPU's Decode stage.}

\end{figure}
The execution flow anatomy of the Decode stage is in \ref{fig:O3_Decode}. Its top-level {\tt tick} function will first sort the instructions ({\tt sortInsts}) by which thread they belong to and then for each thread invoke {\tt checkSignalsAndUpdate}---denoted as simply {\tt checkSignal} in the figures---to read stall and squash signals from the Commit stage. Other signals are categorised as {\tt signals} at this layer. It then squashes or marks the current status accordingly. Afterwards, the core {\tt decode} function is called for the current thread. \profilingRefs{O3-decode-tick} shows the {\tt tick} function's execution time division. We observe that most applications spend at least a third of the decode-tick time inside the core {\tt decode} function, where notable exceptions include {\tt cc\_raw} of GAPBS running 16 O3 cores, {\tt canneal}, {\tt facesim} of PARSEC running $1$ O3 core, {\tt bwaves} and {\tt lbm} of PARSEC, who all spend more time on {\tt checkSignal}.
\profilingFigs{O3CPU}{O3-decode-tick}{Execution time division for Decode's tick function}
Inside the {\tt core} function, if the current thread is \emph{Unblocking}, it invokes {\tt decodeInsts}. \profilingRefs{O3-decode-decode} show its execution time division, where {\tt decodeInsts} constitutes the vast majority of the execution time for all applications.
\profilingFigs{O3CPU}{O3-decode-decode}{Execution time division of {\tt decode}}
\profilingRefs{O3-decodeInsts} show the execution time division for {\tt decodeInsts}, which will check the instructions that are not squashed---mark ready ones and push valid ones to the Rename stage. Furthermore, it resolves branch prediction and squashes mispredictions. Instructions that cannot be processed are saved to the {\tt skidBuffer.}
\profilingFigs{O3CPU}{O3-decodeInsts}{Execution time division for {\tt decodeInsts}}
{\tt checkSignalsAndUpdate} will squash an instruction if such a signal is received from the Commit stage. If we are stalled, the function blocks the stage and stores the instruction for later, and update its own state accordingly. \profilingRefs{O3-decode-Signals} show its execution time division. {\tt Block} tends to be the most time-consuming portion of the function. The second-most prominent portion is spent inside reading from or sending signals to the other stages. {\tt squash} is not prominent for any application or configuration for the current experiment setting.
\profilingFigs{O3CPU}{O3-decode-Signals}{Time division of Decode's {\tt checkSignalsAndUpdate}}

\section{Rename}
\label{sec:o3Rename}
\begin{figure}[H]
  \centering
      \rotatebox{90}{
\includegraphics[width=1.3\textwidth]{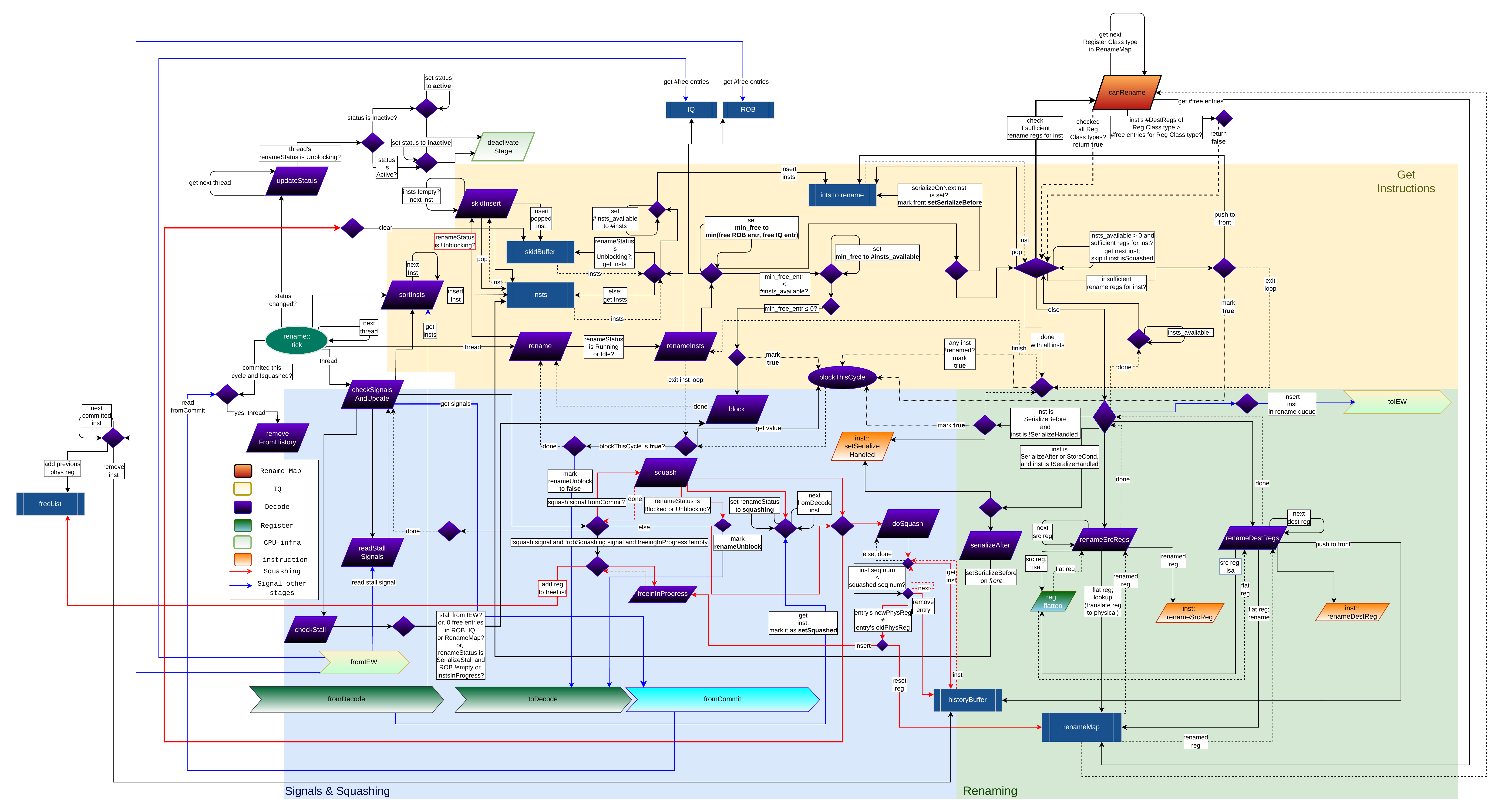}
}
\caption{\label{fig:O3_Rename} Anatomy figure showing the execution flow of the O3 CPU's Rename stage.}
\end{figure}
Figure \ref{fig:O3_Rename} shows the execution flow of the Rename Stage. The stage's top-level {\tt tick} function will begin by sorting the instructions ({\tt sortInsts}) by which thread they belong to. Then for each thread, it checks for a status change using {\tt checkSignalsAndUpdate} ({\tt checkSignal}) and executes the stage's core function {\tt rename}. Afterwards it will for each thread, remove committed instructions' rename history in order to free their register mappings---denoted as {\tt remove-Insts}. These freed registers are added to the {\tt freeList}, which will add them back into the {\tt renameMap} during internal O3 CPU operations. The {\tt renameMap} maintains register mappings and available registers, and is accessed throughout various O3 CPU stages, including the \emph{Rename stage}.

\profilingRefs{O3-rename-tick} show the execution time division at this function using the aforementioned annotations. We can clearly observe that most time is spent inside the core {\tt rename} function and that {\tt checkSignal} is generally the second-most time consuming part. 
\profilingFigs{O3CPU}{O3-rename-tick}{Execution time division for Rename's tick}
If we \emph{flatten} the execution time division, \profilingRefs{O3-rename-flatten} reveal that the bulk of the execution time is spent on {\tt renameInsts} and accessing the instruction's source ({\tt renameSrcRegs}) and destination registers ({\tt renameDestRegs}). This function will read from the IEW stage to see whether the number of free entries in the instruction queue (IQ) or the reordering buffer (ROB), are less than the instructions we have obtained inside the function. We set the variable {\tt insts\_available} to the following:\\
{\tt insts\_available} $=$ \texttt{minimum}(\#instructions, \texttt{minimum}(\#free IQ entries, \#free ROB entries)).

The purpose of this is to compute the maximum number of instructions that can currently be processed.
\profilingFigs{O3CPU}{O3-rename-flatten}{Flattened execution time for Rename's tick}
\profilingRefs{O3-renameInsts} show the execution time division of {\tt renameInsts}, called by the core {\tt rename} function. Likewise, this also supports the notion that most time is spent on accessing source or destination registers, and updating the {\tt RenameMap}. Various instruction operations ({\tt Inst-op}) such as updating their stored registers and other signals is consistently also a contender for being the third-most time-consuming part of the function. Notable exceptions include {\tt facesim} of PARSEC and {\tt lbm} of SPEC, which has a more evenly partitioned execution time. This can also be observed in various GAPBS applications such as {\tt pr\_raw}, though less deviating from the general pattern than {\tt facesim} and {\tt lbm}.
\profilingFigs{O3CPU}{O3-renameInsts}{Execution time division of {\tt renameInsts}}
As previously mentioned, the Rename tick function will initially sort the instructions with {\tt sortInsts}---\profilingRefs{O3-rename-sortInsts} show the execution time division of this function. This function is clearly closer to lower-level C++ operations as {\tt C++-RT} takes up a significant execution time portion, even being the largest time-consumer of many PARSEC applications. Otherwise, the function spends the most time on {\tt signals}---communicating with the interfaces to the other stages.
\profilingFigs{O3CPU}{O3-rename-sortInsts}{Execution time division of Rename's {\tt sortInsts}}
\profilingRefs{O3-rename-Signals} show the execution time division of Rename's {\tt checkSignalsAndUpdate} -- we observe that the overarching tendency is that most time spent is {\tt checkStall}, which will read signals and check whether we have a stalled instruction. However, several GAPBS applications spend comparatively more time on {\tt squash}. This includes {\tt bc\_raw}, {\tt cc\_sv\_raw} and {\tt tc\_raw}. The relative prominence of {\tt squash} is largely application-dependent and partially configuration-dependent, such as in {\tt cc\_raw} where using $16$ cores or $4$ cores and $3$GB of memory decreases {\tt squash}'s prominence.
\profilingFigs{O3CPU}{O3-rename-Signals}{Time division of Rename's {\tt checkSignalsAndUpdate}}

\section{IEW}
\label{res_o3_iew}
\label{sec:o3IEW}
\begin{figure}[H]
  \centering
      \rotatebox{90}{
\includegraphics[width=1.3\textwidth]{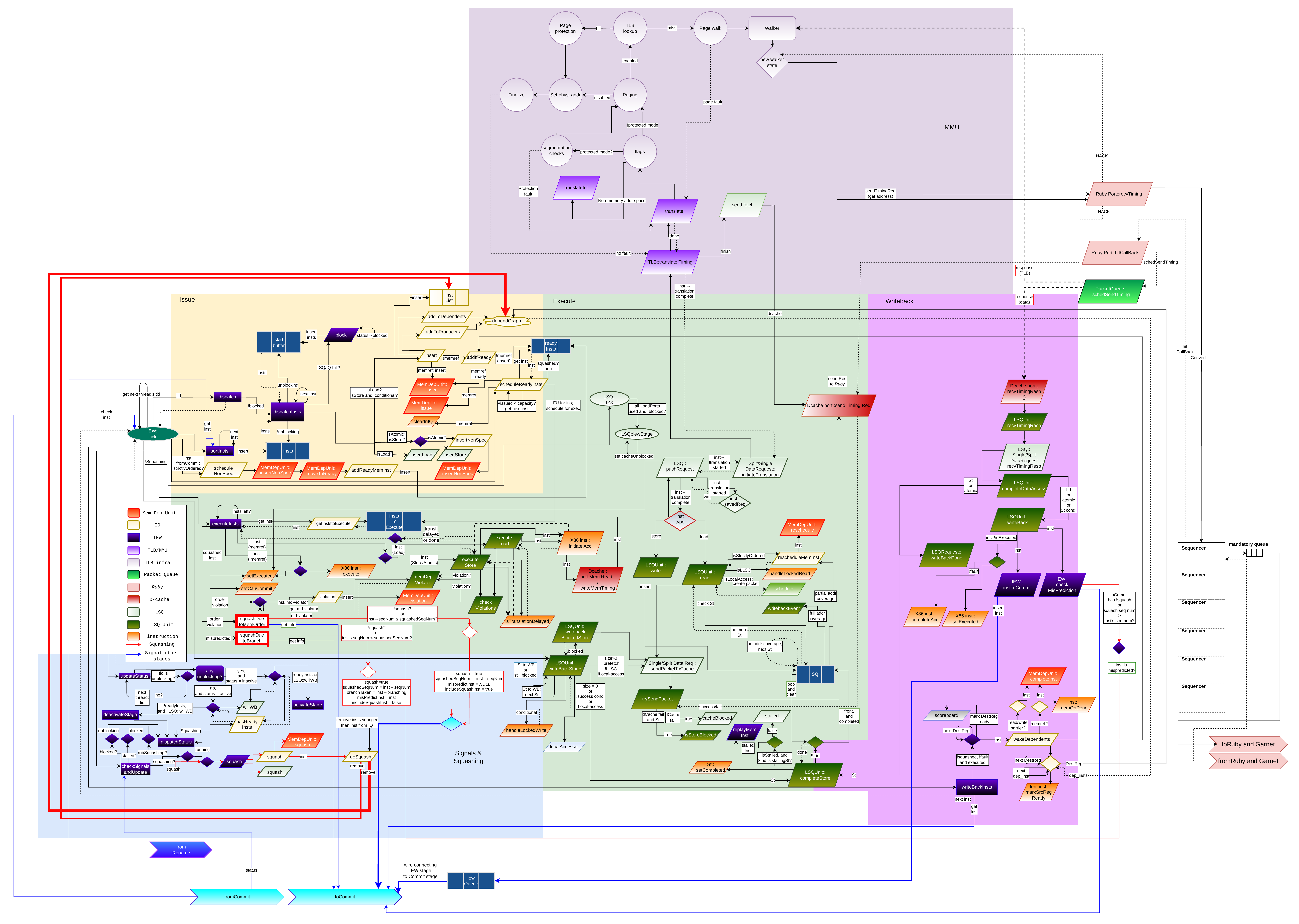}
}
\caption{\label{fig:O3_IEW} Anatomy figure showing the execution flow of the O3 CPU's IEW stage.}
\end{figure}
Figure \ref{fig:O3_IEW} shows the IEW stage's execution flow. Each phase of the IEW {\tt tick} pipelines instructions through the stage. The IEW {\tt tick} will begin by run its {\tt checkSignalsAndUpdate} ({\tt checkSignal}) and then {\tt dispatch} instructions to the instruction queue ({\tt IQ}), potentially updating the stage status ({\tt IEW-updateStatus}) if \textit{e.g.} the {\tt IQ} is full. Afterwards, it will perform the execute ({\tt executeInsts}) and writeback ({\tt writeBackInsts}) stages.

{\tt IEW::dispatch} moves instructions from the \emph{rename} stage to the {\tt IQ} and load/store queue ({\tt LSQ}) if it is a memory operation. The {\tt IQ} will issue instructions once they are ready to execute.

\profilingRefs{O3-IEW-OA} show the overall execution time division of the IEW tick function, whereas \profilingRefs{O3-IEW} shows a more fine-grained division at the same level. We clearly observe that executing instructions ({\tt IEW-exec-insts}) generally makes up the bulk of the execution time. Notable exceptions include the 16 core configurations of {\tt cc\_raw} of GAPBS, {\tt lbm} of SPEC, and {\tt facesim} of PARSEC, which dedicate most of its time to {\tt IEW-updateStatus}. 
\profilingFigs{O3CPU}{O3-IEW-OA}{Division of the overall execution of IEW {\tt tick}}
\profilingFigs{O3CPU}{O3-IEW}{Detailed execution time division of IEW {\tt tick}}
\subsection{Issue}
As mentioned in the prior section, the {\tt dispatch} function handles the IEW stage's \emph{issue} operation. \profilingRefs{O3-IEW-dispatch-flatten} show that this phase spends most of its time on inserting the instructions ({\tt dispatchInsts}) into internal data structures. The execution time is flattened; most of the runtime is spent on updating the instruction queue by inserting valid instructions ({\tt InstructionQueue::insert}) and adding them to the memory dependency unit ({\tt MemDepUnit::insert}).
\profilingFigs{O3CPU}{O3-IEW-dispatch-flatten}{Flattened execution time division of {\tt dispatch}}
\subsection{Execute}
We clearly see in \profilingRefs{O3-exec-Insts} that Loads and Stores constitute the majority of the execution time. This is the execution division of {\tt executeInsts}, in charge of the \emph{execute} phase of the IEW stage. We also observe that the second-most time-consuming operations are related to the dynamic instruction ({\tt D-inst})---including setting its flags such as {\tt setExecuted}.
\profilingFigs{O3CPU}{O3-exec-Insts}{Division of executing instructions}

There are four control paths throughout the IEW stage for executing instructions: (1) Non-memory/compute instructions, (2) load instructions, (3) store instructions and (4) atomic instructions.
\begin{enumerate}
\item \textbf{Non-memory/compute instructions (IEW[speculative execution]$\rightarrow$Commit).} The instruction's {\tt execute} function ({\tt executeInsts}) is invoked and the instruction is subsequently sent to commit. Inside {\tt writeBackInsts}, the IQ's {\tt wakeDependents} is called, where dependent instructions will be added to the list of ready instructions.

\item \textbf{Load instructions (IEW[speculative execution]$\rightarrow$Ruby$\rightarrow$Commit).} {\tt executeLoad} is called by {\tt executeInsts}, which will in turn invoke the instruction's {\tt initiateAcc} (initiate memory access) function, which is its execution context interface. This will invoke {\tt intitiateMemRead} which calls upon the LSQ's {\tt pushRequest} function. It will initiate the address translation via the TLB. Once finished, the function will invoke {\tt read} of the LSQUnit. {\tt read} checks for aliasing with any prior Store instruction---if it can forward, a {\tt writeBackEvent} is scheduled for the next cycle; if the instruction is aliased but cannot forward, the IQ's {\tt rescheduleMemInst} is invoked to cancel the Load instruction. If either of these do not occur, {\tt read} sends packets to the cache via {\tt trySendPacket}. Once the instruction's memory request has been handled (via Ruby), the LSQUnit's {\tt completeDataAccess} will call the {\tt writeBack} function, which in turn invokes the instruction's {\tt completeAcc} to write the instruction's loaded value to the destination register. The Load is then inserted in the Commit queue to be handled in the subsequent Commit stage. {\tt writeBackInsts} will finally mark it as finished and wake up its dependents.

\item \textbf{Store instructions (IEW[speculative execution]$\rightarrow$Commit$\rightarrow$IEW[non-speculative memory write]$\rightarrow$Ruby).}  {\tt executeInsts} will invoke {\tt executeStore}, which in turn triggers the instruction's {\tt initiateAcc}, which will push a request and call the LSQUnit's {\tt write} function. {\tt write} will insert the instruction into the store queue ({\tt SQ}). Once the store instruction has been committed, the LSQUnit's {\tt writeBackStores} is invoked in the IEW stage, which will send the package to the D-cache---the same procedure as with a Load instruction. Once the request's response is handled, {\tt completeDataAccess} will invoke {\tt completeStore} in order to remove the instruction from the SQ.

\item \textbf{Atomic instructions (IEW[stall]$\rightarrow$Commit[ROB flush]$\rightarrow$IEW[non-speculative memory read and write]$\rightarrow$Ruby$\rightarrow$Commit).} A similiar execution flow as store instructions, however, atomic instructions are executed without speculation in the gem5 x86\_64 ISA. This means that {\tt dispatchInsts} will invoke {\tt insertNonSpec} to mark it as such. Once the Commit stage has handled it, the instruction can be scheduled for execution back in the IEW stage, which it does with the {\tt scheduleNonSpec} function. {\tt executeInsts} will call {\tt executeStore}, which in turn invokes the instruction's {\tt initiateAcc}, which will push the request and call {\tt write}. Once {\tt writeBackStores} is invoked, the atomic instruction's request is packaged and sent to the D-cache. Just like the case with a non-atomic store instruction, this will invoke {\tt completeStore} to finalize the instruction's execution. 
\end{enumerate}

A memory instruction's {\tt init acc} invoked in order to initiate its execution, as previously mentioned. \profilingRefs{O3-initacc} show the execution time division at this level in terms of which instruction type's {\tt init acc} was invoked---exposing each applications X86 instruction set. We observe that {\tt Load}s tends to dominate the execution time of most applications. Hence, we examine a {\tt Load} instruction in greater detail.

\profilingFigs{O3CPU}{O3-initacc}{Execution time division of the memref instruction set}
The instruction {\tt init acc} will in turn invoke {\tt pushRequest} of the LSQ to proceed. \profilingRefs{O3-execLoad-OA} shows the \emph{flattened} execution time division at this level for a Load, clearly demonstrating that most time is spent on sending a memory request to \emph{Ruby} or the TLB. Store-dominated applications such as {\tt facesim} of PARSEC spends most execution time inside the TLB at this layer and less time overall since other applications are comparatively much more Load-dominated.
\profilingFigs{O3CPU}{O3-execLoad-OA}{Overall execution time division of executing a {\tt Load}}
\profilingRefs{O3-execLoad} shows a more fine-grained execution time division at the same level.
\profilingFigs{O3CPU}{O3-execLoad}{Execution time division of executing a {\tt Load}}
\subsection{Writeback}
The writeback phase of the IEW stage is triggered at two points: (1) the data cache (D-cache) receives a response from \emph{Ruby}, which will be routed to the LSQUnit's {\tt completeDataAccess} function, completing the memory operation's execution. (2) the top-level {\tt tick} will call {\tt writebackStores} to writeback stores using leftover bandwidth.

\profilingRefs{O3-completeDataAccess} show its execution time division for {\tt completeDataAccess}. We observe how most execution time at this level is spent inside the {\tt writeback} function of the LSQUnit---it will invoke various functions to ensure no mispredictions and faults occurred and if so, proceed with sending the instruction to the Commit stage.
\profilingFigs{O3CPU}{O3-completeDataAccess}{Execution time division for {\tt completeDataAccess}}
We show the execution time division for the {\tt writeback} function in \profilingRefs{O3-writeback}. Across the board, most time is spent inside the dynamic instruction's {\tt completeAcc} procedure. {\tt checkMisprediction} consistently appears in second place.
\profilingFigs{O3CPU}{O3-writeback}{Execution time division for {\tt writeback}}
Afterwards, if the instruction is a Store or Atomic, {\tt completeDataAccess} invokes {\tt completeStore}. This function will mark the store as completed and clear it from the store queue ({\tt SQ}), implemented using a {\tt Circular-Queue} data structure. Additionally, it checks if it has been stalled or not; if it has, then the instruction is marked to be replayed later. \profilingRefs{O3-completeStore} show its execution time division---clearly it spends most of its time on the {\tt SQ}.
\profilingFigs{O3CPU}{O3-completeStore}{Execution time division for {\tt completeStore}}
\profilingRefs{O3-writebackStores} show the execution time division of {\tt writebackStores}, called at the top-level IEW {\tt tick} function---it will execute any leftover ready stores using the available bandwidth. We see that most time is spent on sending a \emph{Ruby} request to the {\tt D-cache}.
  \profilingFigs{O3CPU}{O3-writebackStores}{Exeuction time of {\tt writebackStores}}  
\subsection{Update status}
\label{sec:IEW_updatestatus}
After the writeback phase, it will run its {\tt updateStatus} function to update the IEW stage's status based on the staus of the other stages. \profilingRefs{O3-updateStatus} show the execution time division of this function. The instruction queue's {\tt hasReadyInsts} function is by far the most time-consuming function at this level.
\profilingFigs{O3CPU}{O3-updateStatus}{Execution time division of {\tt updateStatus}}
\profilingRefs{O3-hasReadyInsts} show the \emph{flattened} execution time division of the IQ's {\tt hasReadyInsts} function
\profilingFigs{O3CPU}{O3-hasReadyInsts}{Flattened time division of {\tt hasReadyInsts}}

\section{Commit}
\label{sec:o3Commit}
\begin{figure}[H]
  \centering
      \rotatebox{90}{
\includegraphics[width=1.3\textwidth]{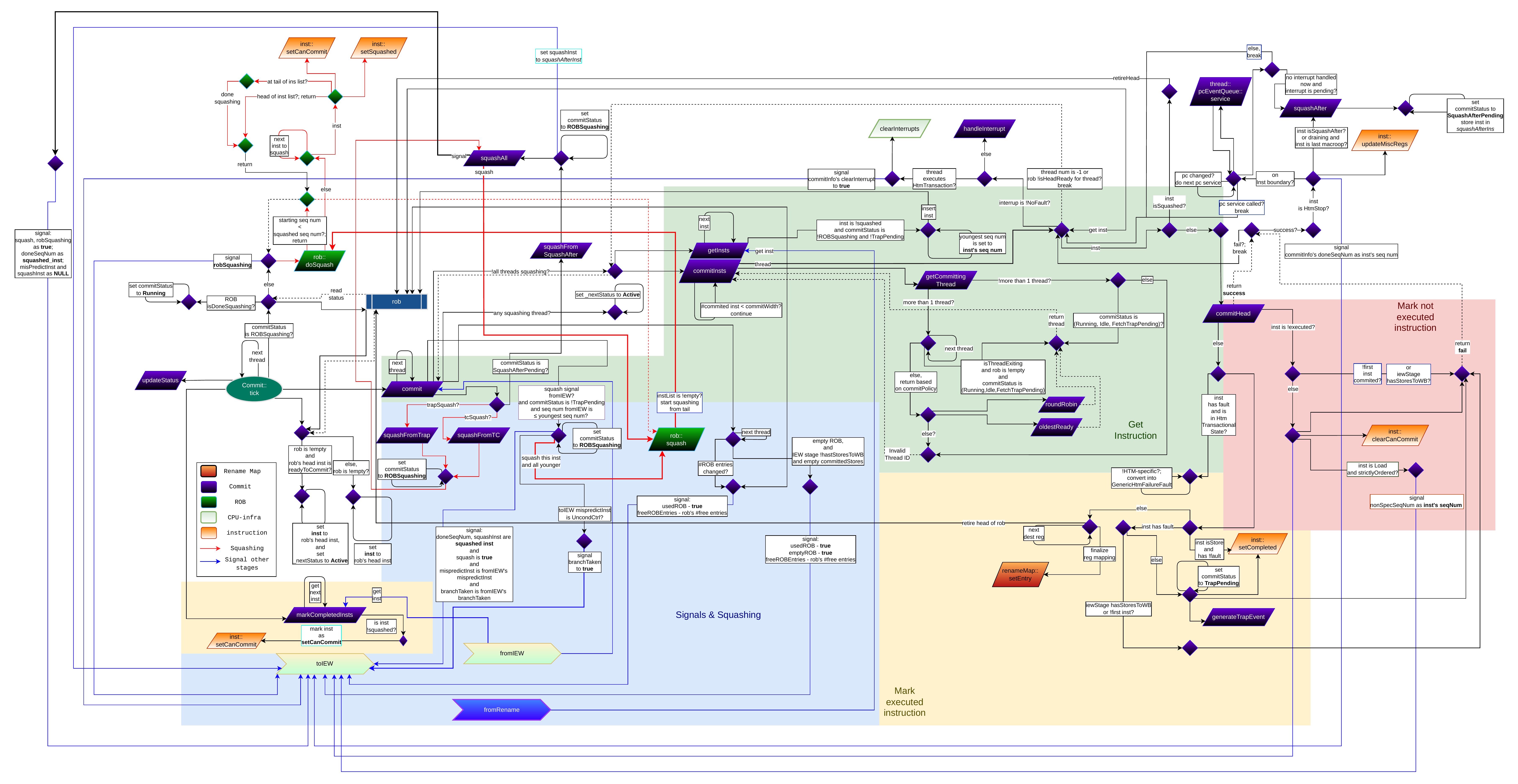}
}
      \caption{\label{fig:O3_Commit} Anatomy figure showing the execution flow of the O3 CPU's Commit stage.}

\end{figure}
The Commit stage's execution flow is detailed in Figure \ref{fig:O3_Commit}. Its top-level {\tt tick} function will call the {\tt commit} function to handle interrupts and squashes and performing the actual instruction committing. Afterwards, it calls {\tt markCompletedInsts} ({\tt markInsts}) to mark instructions received from IEW as ready to be committed for the next tick. It will then check for each thread, if the ROB has more instructions that can commit and set its next status as \emph{active}. It then finishes by calling {\tt updateStatus} to update the CPU and stage status.
\profilingRefs{O3-commit-tick} show the {\tt tick} function's execution time division. Clearly, {\tt commit} consumes the vast majority of the stage's execution time.
\profilingFigs{O3CPU}{O3-commit-tick}{Execution time division for Commit's {\tt tick}}
The {\tt commit} function handles squashes and interrupts. It will commit the instructions with {\tt commitInsts}, fetching them from the ROB with {\tt getInsts}, with the ROB itself getting instructions from the rename stage. It then sends signals to the IEW stage indicating whether we updated the ROB or not.
\profilingRefs{O3-commit-commit} show that {\tt commit} spends the vast majority of its execution time on committing instructions ({\tt commitInsts}) and proportionally much less time updating the ROB and sending signals to other stages.
\profilingFigs{O3CPU}{O3-commit-commit}{Execution time division for {\tt commit}}{\tt commitInsts} will commit as many instructions as possible from the ROB each cycle. For each committing thread, if the head is squashed, it will be retired but not committed. If the ROB's head is ready, it calls {\tt commitHead} to---as the name suggests---commit the head instruction from the ROB. It also checks for any pending interrupts.
\profilingRefs{O3-commitInsts} show the execution time division for {\tt commitInsts}. All applications clearly spend most time inside {\tt commitHead}. The remaining execution time is related to updating the instruction or other infrastructure functions.
\profilingFigs{O3CPU}{O3-commitInsts}{Execution time division for {\tt commitInsts}}
The actual instruction committing is done inside {\tt commitHead}, which commits the head instruction of the ROB. It will check if the instruction has not executed, or is stalled and signal the IEW stage accordingly. Otherwise, if it has no faults we update the {\tt rename map}'s register entries related to the instruction. It will then update internal commit statistics ({\tt Com-Stats}).

\profilingRefs{O3-commitHead} show that {\tt commitHead} spends most of its execution time on updating the commit statistics. The remainder of the execution time is spent on accessing the ROB and marking instructions ({\tt Inst-op}) as \emph{setCompleted} if they were committed as stores, or as \emph{clearCanCommit} if the instruction was not executed (\textit{e.g.} for an atomic instruction, which can only be executed \textit{after} it has been committed).
\profilingFigs{O3CPU}{O3-commitHead}{Execution time division for {\tt commitHead}}

\section{\emph{Ruby}}
\label{res_o3_ruby}
\emph{Ruby} is invoked for each of the O3 CPU's memory accesses. It is used in the Fetch stage (Section \ref{sec:o3Fetch}) and the IEW stage (Section \ref{sec:o3IEW}), for fetching and executing instructions, respectively. We refer to Section \ref{TS_res_ruby} of the TS CPU results for more in-depth explanation of the internal \emph{Ruby} and L1 cache controller functions, as Ruby behaves the same in TS and O3 CPU models.

We show the top-level execution time division in \profilingRefs{O3-ruby}. Applications belong in one of two categories: (1) most time is spent on Garnet or (2) most time is spent inside the L1 cache. All GAPBS applications, {\tt bodytrack}, {\tt canneal}, {\tt dedup}, {\tt facesim}, {\tt ferret} and {\tt fluidanimate} of PARSEC, and {\tt bwaves}, {\tt cactuBSSN}, {\tt gcc}, {\tt imagick}, {\tt lbm}, {\tt omnetpp}, {\tt perlbench}, {\tt x264}, and {\tt xalancbmk} of SPEC---fall into category (1) and are therefore prone to frequently missing in the L1 cache. Conversely, the remaining applications of category (2) tend to hit in the L1 cache.
\profilingFigs{O3CPU}{O3-ruby}{Division execution time of \emph{Ruby}}
\profilingRefs{O3-L1Cache} show that the most time-consuming operation of the L1 cache is the {\tt TransWorker}; executing the current state of the L1 cache controller (CC).
\profilingFigs{O3CPU}{O3-L1Cache}{Execution time division of L1 CC's state managing}
\profilingRefs{O3-L1CC} show the execution time division at the layer of the {\tt TransWorker} function. This layer shows the L1 cache behaviour of the applications---more time spent on {\tt hit} actions indicates they are prone to hit on each respective request. This can be cross-referenced with the relative time spent on Garnet; if an application has comparatively little activity in Garnet and moreso in the L1 cache, it suggest its memory access behaviour causes more L1 cache hits and vice versa.

We observe that it will be largely application-dependent where most execution time is spent. {\tt ifetch-hit} denotes the action performed whenever the L1 cache hits on an instruction fetch. We see this action taking up more of the run-time on applications such as {\tt cc\_raw}, {\tt cc\_sv\_raw} of GAPBS and {\tt freqmine} of PARSEC. Certain applications spend the execution time on other L1 CC operations ({\tt Other-LCC}) -- notably GAPBS applications such as {\tt bc\_raw}, {\tt bfs\_raw}, {\tt pr\_raw} and {\tt pr\_spmv\_raw}.

{\tt load-hit} is triggered when a Load request hits in the L1 cache. Similarly, {\tt store-hit} is whenever a Store request hits in the L1 cache. These are the most time-consuming actions for most PARSEC and SPEC applications.

The TS CPU (Section \ref{TS_res_ruby}) had most of its L1 cache execution time being spent on {\tt ifetch-hit}, whereas the O3 CPU shows a more even distribution. This indicates that the O3 CPU spends proportionally less \emph{Ruby} execution time on instruction fetch and moreso on internal core mechanism for instruction execution.
\profilingFigs{O3CPU}{O3-L1CC}{Execution time division of L1 CC state operations}
\profilingRefs{O3-L1State} show the L1 CC's actions and their execution time division. We explain the prominent actions in Section \ref{TS_res_ruby}. This is the same layer as the previously discussed {\tt hit} actions, but here we show each L1 CC action. We expose what they spend time on aside from the {\tt hit} actions, and it is mostly dedicated to {\tt jj\_sendExclusiveUnblock}---send ``unblock'' to the L2 cache and {\tt k\_popMandatoryQueue}---popping the mandatory queue.
\profilingFigs{O3CPU}{O3-L1State}{Execution time division of the L1 CC's action states}
\profilingRefs{O3-L1-RHC-OA} show the overall execution time of \emph{Ruby}'s hitback call---all applications spend most of their runtime on sending the response packet in the callback function.
\profilingFigs{O3CPU}{O3-L1-RHC-OA}{Overall execution time of the hitback call}
\profilingRefs{O3-L1-RHC} show the execution time division at the same layer but more fine-grained. We expose here that it is specifically the schedule \emph{timing} response function calls that consume most of the execution time.
\profilingFigs{O3CPU}{O3-L1-RHC}{Execution time division of \emph{Ruby}'s hitback call}
\profilingRefs{O3-sendResp} show the execution time division of \emph{Ruby} scheduling a response to the requester. These results also reveal which applications are prone to frequent TLB misses. For example, {\tt bfs\_raw} has many TLB response packets, echoing the TS CPU results in Section \ref{sec:TS_TLB} (\profilingRefs{TS-TLB-Walker}). More time spent on sending a TLB request response also correlates to more proportional time inside \emph{Ruby} at the top-level layer of \profilingRefs{O3-overall}.
\profilingFigs{O3CPU}{O3-sendResp}{Execution time division of \emph{Ruby}'s response}

\section{Garnet}
\label{res_o3_garnet}
For a brief overview of each category, we refer to Section \ref{TS_res_garnet}'s descriptions since they are equally applicable to this section.

We show the execution time division of the Garnet router in \profilingRefs{O3-Garnet-Router}. There is no identifiable performance bottleneck at this layer. However, we do reveal which applications (in relation to each other) tend to miss in the L1 cache, since this is the scenario when Garnet is invoked. This layer shows the router determing the path the request will take through the memory hierarchy.
\profilingFigs{O3CPU}{O3-Garnet-Router}{Execution time division of the Garnet router}
We can see the execution time division for Garnet's network interface (NI) in \profilingRefs{O3-Garnet-NI}. We cannot identify any particular bottleneck operation here either. More time spent on the Garnet router clearly correlates to the time spent on the network interface.
\profilingFigs{O3CPU}{O3-Garnet-NI}{Execution time of the Garnet's network interface}

\chapter{Conclusions}
\label{cha:conclusion}
We illustrated the execution flow of fetching, executing, and committing instructions in each gem5 CPU model, and how execution time is divided across stages. For each benchmark---GAPBS, PARSEC-3.0, and SPEC2017---we profiled gem5 execution time under varying core counts and memory sizes.

For the AS CPU, interaction with \textsf{Ruby} proceeds as a chain of function calls, completed by returning through the chain in reverse. The TS and O3 CPUs are more loosely coupled: they issue a \textsf{timed} request, which enters \textsf{Ruby}, routed via \textsf{Garnet}, and eventually returns a scheduled response. This applies to every memory access: even in the best case---an L1 hit---the request still incurs \textsf{Ruby}'s request-response overhead.

\textbf{AtomicSimple CPU.} gem5 runtime is dominated by instruction fetch. Further, increasing memory sizes forces the simulated OS to manage a larger memory space (with more complex page tables), inflating fetch time, all making the AS CPU bottlenecked by \textsf{Ruby}.  

\textbf{TimingSimple CPU.} For the TS CPU, \textsf{Ruby} as a whole is also the limiting factor. Most runtime inside \textsf{Ruby} is spent on L1 cache controller for instruction fetches (most often a hit in the L1). In Garnet, no single operation dominates gem5 runtime.

\textbf{O3 CPU.} Unlike the TS CPU, the O3 CPU shows a more balanced distribution between its internal pipeline stages and the \textsf{Ruby} memory system. In particular, the IEW stage consumes a large share of runtime, with another significant share spent on \textsf{Ruby} requests for loads and stores.

\bibliographystyle{plainnat}
\bibliography{main}

\end{document}